\documentclass{amsart}
\usepackage{amsmath}
\usepackage{amssymb}
\usepackage{amscd}
\usepackage{amsthm}
\usepackage{epsfig}
\newif\ifmargin
\margintrue
\newif\ifskip
\skiptrue
\newtheorem{oproblem}{Problem}

\newtheorem{theorem}{Theorem}[section]
\newtheorem{lemma}[theorem]{Lemma}
\newtheorem{proposition}[theorem]{Proposition}
\newtheorem{corollary}[theorem]{Corollary}
\newtheorem{definition}[theorem]{Definition}
\newtheorem{remark}[theorem]{Remark}
\newtheorem{example}[theorem]{Example}

\newcommand{\N}{{\mathbb N}}
\newcommand{\Z}{{\mathbb Z}}
\newcommand{\ind}{{\mathcal I}}
\newcommand{\ring}{{\mathcal R}}

\ifskip 
\else

\newcommand{\Z}{{\mathbb Z}}

\newcommand{\card}{{\rm card\ }}

\newcommand{\bl}{\begin{Lemm}}
\newcommand{\be}{\begin{equation}}
\newcommand{\el}{\end{Lemm}}
\newcommand{\ee}{\end{equation}}
\newcommand{\bt}{\begin{Theo}}
\newcommand{\et}{\end{Theo}}
\newcommand{\bp}{\begin{Prop}}
\newcommand{\ep}{\end{Prop}}

\newcommand{\bc}{\begin{Cor}}
\newcommand{\ec}{\end{Cor}}

\fi 
\newenvironment{Proof}{\noindent{\bf Proof:}\ \newline}{\hfill Q.E.D.\newline}
\newcommand{\defeq}{=}

\newcommand{\SOL}{\mathbf{SOL}}
\newcommand{\FPL}{\mathbf{FPL}}

\newcommand{\MSOL}{\mathbf{MSOL}}

\newcommand{\VAR}{\mathbf{VAR}}
\newcommand{\VARFO}{\mathbf{VAR}_1}
\newcommand{\VARSO}{\mathbf{VAR}_2}

\newcommand{\free}{\textit{free}}
\newcommand{\tv}{\mathrm{tv}}
\newcommand{\enabled}{enabled}
\newcommand{\VALORD}{\mathrm{VALORD}}
\newcommand{\rec}{\mathrm{rec}}

\newenvironment{renumerate}{\begin{enumerate}}{\end{enumerate}}

\newcommand{\card}[3]{card_{\mathcal{#1},\bar{#2}}(#3(\bar{#2}))}
\newcommand{\angl}[1]{\langle#1\rangle}
\newif\ifshort
\shorttrue

\begin{document}
\begin{titlepage}
\title[Subset Expansions of Graph Polynomials]{Graph Polynomials:\\
From Recursive Definitions \\ To Subset Expansion Formulas}
\author[B. Godlin]{B. Godlin$^{**}$}
\author[E. Katz]{E. Katz$^{**}$}
\author[J.A. Makowsky]{J.A. Makowsky$^{*}$}

\thanks{$^{*}$
Partially supported by a Grant of the Fund for Promotion
of Research of the Technion--Israel Institute of Technology}
\thanks{$^{**}$
Partially supported by a grant of the Graduate School of the Technion--Israel Institute of Technology}
\email[B. Godlin]{bgodlin@cs.technion.ac.il}
\email[E. Katz]{emika@cs.technion.ac.il}
\email[J.A. Makowsky]{janos@cs.technion.ac.il}
\date{Last revised, December, 7, 2008}

\address{
Department of Computer Science,
\newline
Technion--Israel Institute of Technology,
\newline
32000 Haifa, Israel
}
\begin{abstract}
Many graph polynomials, such as the Tutte polynomial, 
the interlace
polynomial and the matching polynomial, 
have both a recursive definition
and a defining subset expansion formula. 
In this paper we present a general, logic-based framework which 
gives a precise meaning to recursive definitions
of graph polynomials. We then prove that in this framework
every recursive definition of a graph polynomial can be converted
into a subset expansion formula.
\end{abstract}
\end{titlepage}
\maketitle
\small
\tableofcontents
\newpage

\ifskip 
\fi

\section{Introduction}
Graph polynomials are functions from the class
of graphs $\mathcal{G}$ into some polynomial ring $\ring$
which are invariant under graph isomorphisms.
In recent years an abundance of graph polynomials
have been studied. Among the most prominent examples we have
the multivariate Tutte polynomial,
\cite{ar:BollobasRiordan99,ar:Sokal2005a},
the interlace polynomial,
\cite{ar:ArratiaBollobasSorkin2004,ar:ArratiaBollobasSorkin2004a,ar:AignervdHolst2004}
which is really the Martin polynomial, cf.  \cite{ar:EllisM98,ar:Courcelle07},
the matching polynomial and its relatives,
\cite{ar:HeilmannLieb72,bk:LovaszPlummer86,bk:GodsilRoyle01},
and the cover polynomial for directed graphs
\cite{ar:ChungGraham95}.
Older graph polynomials, treated in monographs such as
\cite{bk:Biggs93,bk:Godsil93,bk:Bollobas99,bk:GodsilRoyle01,bk:Diestel05},
are the
characteristic polynomial, 
\cite{bk:CvetkovicDoobSachs1995},
the chromatic polynomial,
\cite{bk:DongKohTeo2005},
and the original Tutte polynomial,
\cite{bk:Bollobas99}.
A general program for the comparative study of graph polynomials
was outlined in \cite{pr:Makowsky06,ar:MakowskyZoo}.

Graph polynomials are usually defined
either recursively or explicitely by a subset expansion formula.
In the case of the polynomial of the Pott's model $Z(G,q,v)$, 
a bivariate graph polynomial
closely related to the Tutte polynomial,
both definitions are easily explained.

Let $G=(V,E)$ be a (multi-)graph. Let $A\subseteq E$ be a
subset of edges. We denote by $k(A)$ the number of connected
components in the spanning subgraph $(V,A)$.
The definition of the Pott's model using a subset expansion
formula is given by
\begin{equation}
Z(G,q,v)=\sum_{A\subseteq E}q^{k(A)}v^{|A|}.
\end{equation}

The general subset expansion formula\footnote{
L. Traldi coined this term in \cite{ar:Traldi04} in the context of the
colored Tutte polynomial.
}
of a graph polynomial
$P(G, \bar{X})$ now takes the form
\begin{equation}
\label{eq:ssexp}
P(G,\bar{X}) =
\sum_{\bar{A}: \angl{G, \bar{A}} \in \mathcal{C}} 
X_1^{f_1(G,\bar{A})} \cdot \ldots \cdot X_{n}^{f_{n}(G, \bar{A})}.
\end{equation}
where $\bar{A}=(A_1, \ldots, A_{\ell})$ 
are relations on  $V(G)$ of arity $\rho(i)$, in other words
$A_i \subseteq V(G)^{\rho(i)}$, 
the summation ranges over
over a family $\mathcal{C}$ of structures of the form $\angl{G,A_1, \ldots, A_{\ell}}$,
and the exponent $f_i(G, \bar{A})$ of the indeterminate $X_i$
is a function from $\mathcal{C}$ into $\N$.
We refer to the right hand side of (\ref{eq:ssexp})
as a {\em subset expansion expression}.

$Z(G,q,v)$ can also be defined recursively.
It satisfies the initial conditions
$Z(E_1)=q$ and
$Z(\emptyset)=1$,
and satisfies a linear recurrence relation 
\begin{eqnarray}
\nonumber Z(G,q,v)&=&v \cdot Z(G_{/e},q,v) + Z(G_{-e},q,v)\\
Z(G_1 \sqcup G_2,q,v)&=&Z(G_1,q,v)\cdot Z(G_2,q,v)
\label{rec-sokal}
\end{eqnarray}
$\sqcup$ denotes the the disjoint union of two graphs,
and for $e \in E$, the graph
$G_{-e}$ is obtained from $G$ by deleting the edge $e$,
and $G_{/e}$ is obtained from $G$ by contracting the edge $e$.
To show that $Z(G,q,v)$ is well-defined using the recurrence relation
\ref{rec-sokal},
one chooses an ordering of the edges and shows that
the resulting polynomial does not depend on the particular
choice of the ordering.

In the case of the Tutte polynomial it is a bit more complicated,
as the recursion involves case distinction depending on whether the
elimitated edge is a bridge, a loop or none of these.
These conditions can be formulated as guards. 

For most prominent graph polynomials, such as the
chromatic polynomial, the Tutte polynomial, the interlace
polynomial, and the cover polynomial for directed graphs,
there exist both a recursive definition using
a linear recurrence relation and a subset expansion
formula. In each case the author proposes the two definitions
and proves their equivalence. 

In this paper we show how to convert a
definition using a linear recurrence relation
into a subset expansion formula.
For this  to make sense we define an appropriate framework.
A special case of
subset expansion formulas is the notion of a graph polynomial
definable in Second Order Logic $\SOL$, introduced
first \cite{ar:MakowskyTARSKI} and further studied in
\cite{ar:MakowskyZoo,ar:KotekMakowskyZilber08}.
The exact definitions are given in Section \ref{se:sol}.
Roughly speaking, $\SOL$-definable graph polynomials
arise when in the subset expansion formula
the class $\mathcal{C}$ is required to be definable
in $\SOL$, and similar conditions are imposed on the exponents
of the indeterminates.

The recursive definition given above relies on the fact
that every graph can be reduced, using edge deletion
and edge contraction, to a set of isolated vertices.
In a last step the isolated vertices are removed one by one.
Using a fixed ordering of the edges and vertices, one can evaluate the
recurrence relation. Finally one has to show that this evaluation
does not depend on the ordering of the edges, provided the that in that ordering
the vertices appear after all the edges.

In general, the two operations, edge deletion and contraction,
will be replaced by a finite set of $\SOL$-definable
transductions $T_1, \ldots, T_{\ell}$, 
which decrease the size of the graph, and
which depend on a fixed number of vertices or edges, the contexts,
rather than just on a single edge.
For certain orderings of the vertices and edges, this
allows us to define a deconstruction tree of the graph $G$.

The recursive definition now takes the form
\begin{equation}
    P(G)\defeq
              \sum_{i\in \{1,\ldots,\ell\} }
              \sigma_i \cdot P(T_i[G,\vec{x}])
\end{equation}
where $\vec{x}$ is the context and $\sigma_i$
are  the coefficients of the recursion.
Furthermore, the recurrence relation is \emph{linear}
in $P(T_i[G,\vec{x}])$.
It can be evaluated using the deconstruction tree.
To assure that this defines a unique graph polynomial
one has to show that the evaluation is independent
of the ordering.
The exact definitions are given in Section \ref{se:rec-def}.

Our main result, Theorem \ref{th:main-result}, 
now states that, indeed, every order
invariant definition of a graph polynomial $P$ 
using a linear recurrence relation can be converted into
a definition of $P$ as a $\SOL$-definable graph polynomial.
It seems that the converse is not true, but we have not been able
to prove this.

In Section \ref{se:application} we discuss a graph polynomial
introduced in \cite{ar:NobleWelsh99}, which is provably not a
$\SOL$-definable graph polynomial.
It is defined by a subset expansion formula, where 
the exponents
$f_i(G, \bar{A})$ depend on $i$, which is not allowed in our definition
of $\SOL$-definable graph polynomials.

The choice of $\SOL$ is rather pragmatic. It makes exposition clear and covers
all the examples from the literature. The logic $\SOL$ could be replaced
by the weaker Fixed Point Logic $\FPL$ or by extensions of $\SOL$, as they are used 
in Finite Model Theory, cf. \cite{bk:FMT}.
The polynomial introduced in \cite{ar:NobleWelsh99} would still be
an example without recursive definition 
as long as the exponents $f_i(G, \bar{A})$ are not allowed to depend on $i$.

The paper is organized as follows.
In Section \ref{se:logic} we collect the background material for Second Order
Logic. In Section \ref{se:sol-gp} we give a rigorous definition
of $\SOL$-definable graph polynomials and collect their basic properties.
In Section \ref{se:rec-def} we present our general framework for
recursive definitions of graph polynomials, and discuss examples in detail.
In Section \ref{se:main} we state and prove our main theorem.
In Section \ref{se:main_examples} we show two derivations of subset expansion
formulas, for the universal edge elimination polynomial and the cover polynomial,
using the technique of the proof
of Theorem \ref{th:main-result}.
These derivations give the subset expansion formulas known in the
literature.
In Section \ref{se:application} we discuss a polynomial which is
given by a subset expansion formula but has no recursive definition
in our sense. Finally, in Section \ref{se:conclu}
we draw conclusions and discuss further research.

\subsubsection*{Acknowledgments}
The authors would like to thank I. Averbouch,
B. Courcelle,
T. Kotek
for valuable discussions and suggestions.

\section{Logic and Translation Schemes}
\label{se:logic}
In this section we give a rather detailed definition
of $\SOL$ and the formalism of translation schemes, 
because the notational technicalities are needed in
our further exposition.

A \emph{vocabulary} $\tau$ is a finite set of relation symbols,
function symbols and constants. It can be many-sorted.
In this paper, we shall only deal with vocabularies which do not
contain any function symbols. $\tau$-structures are interpretations
of vocabularies. Sorts are mapped into non-empty sets -
the sort universes. Relation symbols are mapped into
relations over the sorts according to their specified
arities. Constant symbols are mapped onto elements
of the corresponding sort-universes.
We denote the set of all $\tau$-structures by $Str(\tau)$.
For a $\tau$-structure $\mathcal{M}$, 
we denote its universe by $A^\mathcal{M}$,
or, in short, $A$, if the $\tau$-structure is clear from the context.
For a logic $\mathcal{L}$, $\mathcal{L}(\tau)$ denotes the
set of $\tau$-formulas in $\mathcal{L}$.

\subsection{Second Order Logic ($\SOL$)}
\label{se:sol}
We denote relation symbols by bold-face letters,
and their interpretation by the corresponding
roman-face letter.
\begin{definition}[Variables]$ $\\
\label{def:SOL-vars}
\begin{renumerate}

\item $v_i$ for each $i \in \N$.
These are {\em individual variables} $(\VARFO)$.

\item
$U_{r,i}$ for each $r,i \in \N, r \geq 1$.
These are {\em relation variables}  $(\VARSO)$.
$r$ is the arity of $U_{r,i}$.

\end{renumerate}
We denote the set of variables by $\VAR$.
\\

Given a {\em non-empty finite} set $A$,
an $A$-\emph{interpretation} is a map
$$
I_A: \VAR \rightarrow  A \cup \bigcup_r \mathbb{P}(A^r)
$$
such that
$I_A(v_i) \in A$ and
$I_A(U_{r,i}) \subseteq A^r$.
\end{definition}
$ $\\

We define term $t$ and formula $\phi$ inductively, and associate with them a set
of first and second-order free variables denoted by $free(t)$, $free(\phi)$ respectively.

\begin{definition}[$\tau$-term]
\label{def:tau-term}
A $\tau$-term is of the form $v$ or $c$ where
$v$ is a variable and $c$ is some constant in $\tau$.
$\free(v) = \{v\},~\free(c) = \emptyset$.
\end{definition}

\begin{definition}[Atomic formulas]
\label{def:SOL-atomic}
$ $\\
{\emph Atomic formulas} are of the form

\begin{renumerate}
\item
$(t_1 \simeq t_2)$ where $t_1, t_2$ are $\tau$-terms, and
$\free(t_1 \simeq t_2) = \free(t_1) \cup \free(t_2)$.

\item
$\phi$ of the form $U_{r,j}(t_1, t_2, \ldots , t_r)$
where $U_{r,j}$ is a relation variable, and $t_1, t_2, \ldots , t_r$ are
$\tau$-terms, and $\free(\phi) = \{U_{r,j}\}\cup \bigcup_{i=1}^r \free(t_i)$.

\item
$\phi$ of the form $R(t_1, t_2, \ldots , t_r)$
where $R\in \tau$ is a relation, and $t_1, t_2, \ldots , t_r$ are
$\tau$-terms, and $\free(\phi) = \bigcup_{i=1}^r \free(t_i)$.

\end{renumerate}
\end{definition}

We now define inductively the set of $SOL$-formulas $\SOL$.
\\

\begin{definition}[SOL formulas] \label{def:SOL-formulas}$ $\\
\begin{renumerate}
\item
Atomic formulas $\phi$ are in $\SOL$ with $\free(\phi)$ as defined before.
\item
If $\phi_1$ and $\phi_2$ are in $\SOL$
then $\phi$ of the form
$(\phi_1 \vee \phi_2)$,
$(\phi_1 \wedge \phi_2)$ or
$(\phi_1 \rightarrow \phi_2)$
is in $\SOL$
with $\free(\phi) = \free(\phi_1) \cup \free(\phi_2)$.
\item
If $\phi_1$ is in $\SOL$
then $\phi = \neg \phi_1$ is in $\SOL$
\\
with $\free(\phi)=\free(\phi_1)$.
\item
If $\phi_1$ is in $\SOL$
then $\phi$ of the form
$\exists v_j \phi$,
$\forall v_j \phi$,
\\
is in $\SOL$ with $\free(\phi) = \free(\phi_1) - \{v_j\}$.
\item
If $\phi_1$ is in $\SOL$
then $\phi$ of the form
$\exists U_{r,j} \phi$ or
$\forall U_{r,j} \phi$
\\
is in $\SOL$ with $\free(\phi) = \free(\phi_1) - \{U_{r,j}\}$.
\end{renumerate}
\end{definition}

$ $\\
\subsection{Translation schemes and deconstruction schemes}
\label{se:transduct}

\begin{definition}[Translation scheme $\Phi$]
\label{def:translation-scheme} Let $\tau = \{Q_1,
\ldots , Q_k\}$ and  $\sigma = \{R_1,
\ldots , R_m\}$ be two vocabularies and $\rho(R_i)$ ($\rho(Q_i)$) be the arity of
$R_i$ ($Q_i$). Let $\mathcal{L}$ be a fragment of $SOL$, such as $FOL$,
$MSOL$, $\exists MSOL$, $FPL$ (Fixed Point Logic), etc.

A tuple of ${\mathcal{L}}(\tau)$ formulae $\Phi = \langle \phi, \psi_1, \ldots, \psi_m \rangle$
such that $\phi$ has exactly one free first order variable and each $\psi_i$ has
$\rho(R_i)$ distinct free first order variables is a $\tau-\sigma$-translation scheme.
\end{definition}
In this paper we use only translation schemes in which $\phi$ has exactly one free variable.
Such translation schemes are called \emph{non-vectorized}.

In our case $\{x: \phi(x)\} \subset A$ holds. Such translation schemes are called \emph{relativized}.

We now define the \emph{transduction} which is the semantic map associated with $\Phi$.

\begin{definition}[The induced transduction $\Phi^{\star}$]
\label{def:transduction}
Given a $\tau-\sigma$-translation scheme $\Phi$, the function
$\Phi^{\star}: Str(\tau) \rightarrow Str(\sigma)$
is a (partial) function from $\tau$-structures to $\sigma$-structures.
$\Phi^{\star}[{\mathcal M}]$ is defined by:
\begin{enumerate}
\item the universe of $\Phi^{\star}[{\mathcal M}]$ is the set
$$A^{\Phi^{\star}[\mathcal{M}]} =\{a \in A : {\mathcal M} \models \phi(a) \}$$
\item the interpretation of $R_i$ in $\Phi^{\star}[{\mathcal M}]$ is the set
$$
R_{i}^{\Phi^{\star}[\mathcal{M}]}
=\{\bar{a} \in {(
A^{\Phi^{\star}[\mathcal{M}]} 
)}^{\rho(R_i)}:
{\mathcal M} \models \psi_i(\bar{a}) \}.
$$
\end{enumerate}
\end{definition}

Next we define the syntactic map associated with $\Phi$, the translation.

\begin{definition}[The induced translation $\Phi^{\sharp}$]
\label{def:translation}
Given a $\tau-\sigma$-translation scheme $\Phi$ we define a function
$\Phi^{\sharp}: {\mathcal L}(\sigma) \rightarrow {\mathcal L}(\tau)$
from ${\mathcal L}(\sigma)$-formulae to ${\mathcal L}(\tau)$-formulae
inductively as follows:
\begin{enumerate}
\item
For $R_i \in \sigma$ with $\rho(R_i)=m$ and $\theta = R_i( x_1,\ldots, x_m)$,
we put
$$\Phi^{\sharp}(\theta) = \left( \psi_i(x_1, \ldots , x_m)
                                 \wedge \bigwedge_{j=1}^m \phi(x_j) \right)
$$
\item
This also works for equality and  relation variables $U$
instead of relation symbols $R$.

\item
For the boolean connectives, the translation
distributes, i.e.
\begin{enumerate}
\item
if $\theta = (\theta_1 \vee \theta_2)$
then
$\Phi^{\sharp}(\theta)=
(\Phi^{\sharp}({\theta_1}) \vee \Phi^{\sharp}({\theta_2}))$
\item
if $\theta = \neg\theta_1$ then
$\Phi^{\sharp}(\theta)= \Phi^{\sharp}(\neg{\theta_1})$
\item
similarly for $\wedge$ and $\rightarrow$.
\end{enumerate}

\item
For the existential quantifier, we use relativization to $\phi$:
\newline
If $\theta = \exists y \theta_1$, we put
$$\Phi^{\sharp}(\theta) = \exists y (\phi(y) \wedge \Phi^{\sharp}(\theta_1)(y)).$$

\item
For the universal quantifier, we also use relativization to $\phi$:
\newline
If $\theta = \forall y \theta_1$, we put
$$\Phi^{\sharp}(\theta ) = \forall y (\phi(y) \rightarrow \Phi^{\sharp}(\theta_1)(y)).$$

\noindent
This concludes the inductive definition for first order logic $FOL$.\\

\item
For second order quantification of variables $V$ of arity $\ell$ and
a vector $\bar{a}$ of length $\ell$ of first order variables or constants,
we translate $\theta=\exists V(\theta_1(V))$ by treating $V$
as a relation symbol above $A$ and put
$$\Phi^{\sharp}(\theta) = \exists V \left(\forall \bar{v} \left[
  ~V(\bar{v}) \rightarrow
     (\bigwedge_{i=1}^\ell \phi(v_i)) \right]
     \wedge \Phi^{\sharp}(\theta_1)(V)~
  \right)$$

\item
For $\theta=\forall V(\theta_1(V)),~\rho(V)=\ell$ the relativization yields:
$$\Phi^{\sharp}(\theta) = \forall V \left( \left[
     \forall \bar{v} (V(\bar{v}) \rightarrow \bigwedge_{i=1}^\ell \phi(v_i)) \right] \rightarrow
     \Phi^{\sharp}(\theta_1)(V)~
  \right)$$
\end{enumerate}
\end{definition}

Next we present the well known fundamental property of translation schemes~\cite{ar:MakowskyTARSKI}.

\begin{theorem}[Fundamental Property]$ $\\
\label{th:trans-funamental}

Let $\Phi = \langle \phi, \psi_1, \ldots, \psi_m \rangle$
be a $(\tau-\sigma)$-translation scheme in a logic $\mathcal L$.
Then the transduction $\Phi^{\star}$
and the translation $\Phi^{\sharp}$ are linked in $\mathcal L$.
In other words, given
${\mathcal M}$ be a $\tau$-structure and
$\theta$ be a ${\mathcal L}(\sigma)$-formula\\
then
$${\mathcal M} \models \Phi^{\sharp}(\theta)~~
\Leftrightarrow
~~\Phi^{\star}({\mathcal M}) \models \theta$$
\end{theorem}

\begin{center}
\begin{figure}
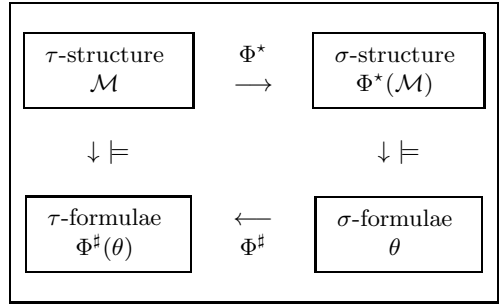

\begin{tabular}{|ccc|}
\hline
&&\\
\fbox{\begin{tabular}{c}
  $\tau$-structure \\
  ${\mathcal M}$
\end{tabular}
}
&
\begin{tabular}{c}
  $\Phi^{\star}$\\
  $\longrightarrow$
\end{tabular}
&
\fbox{\begin{tabular}{c}
  $\sigma$-structure \\
  $\Phi^{\star}({\mathcal M})$
\end{tabular}
}\\
&&\\
$\downarrow~\models$ & & $\downarrow~\models$ \\
&&\\
\fbox{\begin{tabular}{c}
  $\tau$-formulae \\
  $\Phi^{\sharp}(\theta)$
\end{tabular}
}
&
\begin{tabular}{c}
  $\longleftarrow$ \\
  $\Phi^{\sharp}$
\end{tabular}
&
\fbox{\begin{tabular}{c}
  $\sigma$-formulae \\
  $\theta$
\end{tabular}
}\\
&&\\
\hline
\end{tabular}
\caption{A diagram of translation scheme $\Phi$}\label{fig:translation-scheme}
\end{figure}
\end{center}

The property is illustrated in Figure~\ref{fig:translation-scheme}.

\begin{proposition}
\cite{ar:MakowskyTARSKI}
Let $\Phi$ be a $\tau-\sigma$-translation scheme which is either in SOL or in MSOL.
\begin{enumerate}
    \item If $\Phi$ is in MSOL and non-vectorized, and $\theta$ is in MSOL then
        $\Phi^{\sharp}(\theta)$ is in MSOL
    \item If $\Phi$ is of quantifier rank $q$ and has $p$ parameters, and $\theta$ is a
        $\sigma$-formula of quantifier rank $r$, then the quantifier rank of $\Phi^{\sharp}(\theta)$
        is bounded by $r+q+p$.
\end{enumerate}
\end{proposition}

\section{$\SOL$-polynomials}
\label{se:sol-gp}

$\SOL$-polynomial expressions are expressions the interpretation of which
are graph polynomials.
We define $\SOL$-polynomial expressions inductively.

\subsection{$\SOL$-polynomial expressions}

Let the domain $\ring$ be a commutative semi-ring, which contains
the semi-ring of the integers $\N$. For our discussion it is sufficient
for $\ring$ to be $\N$, $\Z$ or polynomials over these, 
but the definitions generalize. 
Our polynomials have a fixed set of indeterminates $\ind$. 
We denote the indeterminates by capital letters $X,Y,\ldots$
We distinguish them from the variables of $\SOL$ which we denote by lowercase letters
$v,u,e,x,\ldots$

%
%

\begin{definition}[$\SOL$-monomial expressions]
\label{def:monomials}
We first define the {\em $\SOL$-monomial expressions}
inductively.
\begin{renumerate}
\item $a\in \ring$
is a $\SOL$-monomial expression,
and $\free(a)=\emptyset$.\\

\item Given a logical formula $\varphi$, $\tv(\varphi)$
is a $\SOL$-monomial expression.\\
$\tv(\varphi)$ stands for the truth value of the formula $\varphi$.
\item
For a finite product $M=\prod_{i=1}^r t_i$ of monomial expressions $t_i$,
$M$ is a $\SOL$-monomial expression, and
$\free(M) = \bigcup_{i=1}^r \free(t_i)$.\\

\item
Let $\phi(\bar{a},\bar{b},\bar{U})$ be a $\tau \cup \{\bar{a},\bar{b},\bar{U}\}$-formula in $\SOL$,
where $\bar{a} = (a_1, \ldots , a_m)$ is a finite sequence of constant
symbols not in $\tau$, $\bar{b}$ is a sequence of free individual variables,
and $\bar{U}$ is a sequence of free relation variables.
Let $t(\bar{a},\bar{b},\bar{U})$ be a $\SOL$-monomial expression. 
Then
$$
M(\bar{b},\bar{U})=\prod_{\bar{a}:
                                  \phi(\bar{a},\bar{b},\bar{U}) }
                              t(\bar{a},\bar{b},\bar{U})
$$
is a $\SOL$-monomial expression and\\
$\free(M)=\free(t)\cup \free(\phi) \setminus \{\bar{a}\}$.
Thus, $\prod$ is a binding operator which binds $\bar{a}$.

\end{renumerate}
\end{definition}

\begin{definition}[$\SOL$-polynomial expressions]
\label{def:polynomials}
The $\SOL$-polynomial expressions are defined inductively:
\begin{renumerate}
\item
$\SOL$-monomial expressions are
$\SOL$-polynomial expressions.\\

\item
For a finite sum $S=\sum_{i=1}^r t_i$ of $\SOL$-polynomial expressions $t_i$,
$S$ is a $\SOL$-polynomial expression, and
$\free(S) = \bigcup_{i=1}^r \free(t_i)$.\\

\item
Let $\phi(\bar{a},\bar{b},\bar{U})$ be a $\tau \cup \{\bar{a},\bar{b},\bar{U}\}$-formula in $\SOL$
where $\bar{a} = (a_1, \ldots , a_m)$ is a finite sequence of constant
symbols not in $\tau$, $\bar{b}$ is a sequence of free individual variables,
and $\bar{U}$ is a sequence of free relation variables.
Let $t(\bar{a},\bar{b},\bar{U})$ be a $\SOL$-polynomial expression. Then
$$
S(\bar{b},\bar{U})=\sum_{\bar{a}:
                           \phi(\bar{a},\bar{b},\bar{U})}
                             t(\bar{a},\bar{b},\bar{U})
$$
is a $\SOL$-polynomial expression and\\
$\free(P)=\free(t)\cup \free(\phi) \setminus \{\bar{a}\}$.
Thus, $\sum$ is a binding operator which binds $\bar{a}$.

\item
Let $\phi(\bar{W},\bar{b},\bar{U})$ be a $\tau \cup \{\bar{W},\bar{b},\bar{U}\}$-formula in $\SOL$
where $\bar{W} = (W_1, \ldots , W_m)$ is a finite sequence of relation
symbols not in $\tau$, $\bar{b}$ is a sequence of free individual variables,
and $\bar{U}$ is a sequence of free relation variables.
Let $t(\bar{W},\bar{b},\bar{U})$ be a $\SOL$-polynomial expression. 
Then
$$
S(\bar{b},\bar{U})=\sum_{\bar{W}:
                           \phi(\bar{W},\bar{b},\bar{U})}
                             t(\bar{W},\bar{b},\bar{U})
$$
is a $\SOL$-polynomial expression and\\
$\free(P)=\free(t)\cup \free(\phi) \setminus \{\bar{W}\}$.
Again, $\sum$ is a binding operator which binds $\bar{W}$.

\end{renumerate}
\end{definition}

%
%
Note that our definition of $\SOL$-polynomial expressions is the normal form
definition as it appears for example in~\cite{ar:KotekMakowskyZilber08}.
We use only the normal form in this paper.

From our definitions the following is obvious.
\begin{proposition}
Every $\SOL$-polynomial expression is also a subset expansion expression,
where $\mathcal{C}$ is $\SOL$-definable.
\end{proposition}

\subsection{Interpretations of $\SOL$-polynomial expressions}$ $\\

Let $G$ be a graph and $z$ be an assignment of variables
to elements of the graph.
The interpretation $e(S, G, z)$ of a $\SOL$-polynomial expression
$S$ will be an element in the polynomial ring $\ring$.
We shall associate with each $\SOL$-polynomial expression $S$
a graph polynomial $S^*$ defined by
$ S^*(G) = e(S,G,z)$.
We shall say that $P(G,\bar{X})$ is a $\SOL$-polynomial
if there is a $\SOL$-polynomial expression $S$ such that
for all graphs $G$ we have $P(G, \bar{X}) = S^*(G)$.

We now proceed with the precise definitions.

\begin{definition}[Variable assignment]$ $\\
\begin{renumerate}
    \item Given a $\tau$-structure $\mathcal{M}$ 
        with domain $A^{\mathcal{M}}$,
        an assignment $z$ is an $A^{\mathcal{M}}$-interpretation of $\VAR$.
    \item We denote the set of all assignments above by $Ass(\mathcal{M})$.
    \item Let $z_1$ and $z_2$ be two assignments in $Ass(\mathcal{M})$.
        Let $v \in \VAR$ be a variable. We write $z_1 =_v z_2$ if for every
        variable $u \neq v$ we have that $z_1(u) = z_2(u)$. 
\end{renumerate}
Our notation naturally extends to vectors of variables.
\end{definition}

\begin{definition}[Interpretation of $\SOL$-monomial expressions]
Given a $\tau$-structure $\mathcal{M}$ and an assignment $z \in Ass(\mathcal{M})$,
the interpretation $e(S, \mathcal{M}, z)$ of a $\SOL$-monomial expression
$S$ is defined as follows:
\label{def:monomials-meaning}
\begin{renumerate}
\item If $S = a\in \ring$, $e(S,\mathcal{M}, z) = a$.\\

\item Given a logical formula $\varphi$,
\[ e(\tv(\varphi), \mathcal{M}, z)\defeq
      \left\{ \begin{array}{ll}
        1^{\ring} & if~~\mathcal{M},z \models\varphi \\
        0^{\ring} & otherwise
      \end{array} \right.
\]

\item
For a finite product $S=\prod_{i=1}^r t_i$ of monomials $t_i$,
$$e(S,\mathcal{M},z) = \prod_{i=1}^r e(t_i, \mathcal{M},z).$$

\item
If $S(\bar{b},\bar{U})=\prod_{\bar{a}:
                                  \phi(\bar{a},\bar{b},\bar{U}) }
                              t(\bar{a},\bar{b},\bar{U})$
then
$$
e(S(\bar{b},\bar{U}),\mathcal{M}, z)=\prod_{
                \begin{array}{c}
                z_1~s.t.~z_1 =_{\bar{a}} z~and \\
                \mathcal{M},z_1 \models \phi(\bar{a},\bar{b},\bar{U})
                \end{array}
                }
                e(t(\bar{a},\bar{b},\bar{U}),\mathcal{M},z_1).
$$
\end{renumerate}
We call the expression $S$ a \emph{short product} as the number of elements
in the product is polynomial in the size of the universe of $\mathcal{M}$.
\end{definition}

The degree of the polynomial $e(S,\mathcal{M},z)$, 
is polynomially bounded by the size of $\mathcal{M}$.

\begin{definition}[Interpretation of $\SOL$-polynomial expressions]
\label{def:polynomials-meaning}
Given a $\tau$-structure $\mathcal{M}$ and an assignment $z \in Ass(\mathcal{M})$,
the meaning function $e(S, \mathcal{M}, z)$ of a $\SOL$-polynomial
expression $S$ is defined as follows:
\begin{renumerate}

\item
For a finite sum $S=\sum_{i=1}^r t_i$ of $\SOL$-polynomial expressions $t_i$,\\
$e(S,\mathcal{M},z) = \sum_{i=1}^r e(t_i,\mathcal{M},z).$\\

\item
If
$
S(\bar{b},\bar{U})=\sum_{\bar{a}:
                           \phi(\bar{a},\bar{b},\bar{U})}
                             t(\bar{a},\bar{b},\bar{U})
$
then
$$
e(S(\bar{b},\bar{U}),\mathcal{M}, z)=\sum_{
                                        \begin{array}{c}
                                            z_1~s.t.~z_1 =_{\bar{a}} z~and \\
                                            \mathcal{M},z_1 \models \phi(\bar{a},\bar{b},\bar{U})
                                        \end{array}
                                            }
                                       e(t(\bar{a},\bar{b},\bar{U}),\mathcal{M},z_1).
$$
We call the expression $S$ a \emph{short sum} as the number of summands
in the sum is polynomially bounded in the size of the universe of $\mathcal{M}$.\\

\item
If
$
S(\bar{b},\bar{U})=\sum_{\bar{W}:
                           \phi(\bar{W},\bar{b},\bar{U})}
                             t(\bar{W},\bar{b},\bar{U})
$
then
$$
e(S(\bar{b},\bar{U}),\mathcal{M}, z)=\sum_{
   \begin{array}{c}
   z_1~s.t.~z_1 =_{\bar{W}} z~and \\
  \mathcal{M},z_1 \models \phi(\bar{W},\bar{b},\bar{U})
 \end{array}
 }
 e(t(\bar{W},\bar{b},\bar{U}),\mathcal{M},z_1).
$$
We call such a sum $S$ a \emph{long sum} as the number of addends
in the sum can be exponential in the size of the universe of $\mathcal{M}$.

\item
A $\SOL$-polynomial expression $S$ is \emph{short}
if it does not contain any long sums as subexpressions.
\end{renumerate}
\end{definition}

With these definition we have
\begin{proposition}
Let $S$ be an
$\SOL$-polynomial expression. 
Let $S^*$ be defined by
$ S^*(G) = e(S,G,z) $. 
Then there is a graph polynomial $P(G,\bar{X})$ 
such that
for all graphs $G$
we have $P(G, \bar{X}) = S^*(G)$.
\end{proposition}
We say that
$P(G,\bar{X})$ 
is a $\SOL$-polynomial
if there is a $\SOL$-polynomial expression $S$  such that
$P(G, \bar{X}) = S^*(G)$.

\subsection{Examples}
\label{se:solpol-examples}
\ \\
In the following section we represent graphs using one of 
the following two vocabularies:
$\tau_{graph(1)} = \{E\}$ and $\tau_{graph(2)} = \{N\}$. 
For vocabulary $\tau_{graph(1)}$, the universe of the graph is
the set of its vertices, 
$A \defeq V$, and $R \defeq E\subseteq V^2$ 
is the relation that represents the edges.
For $\tau_{graph(2)}$, the universe consists of both vertices and edges, 
$A \defeq V \cup E$, and
$R \defeq N \subseteq V \times E$ relates vertices to adjacent edges.

Below are some formulas we need for many of the examples below.
All the formulas are in $\SOL(\tau_{\mathrm{graph}(1)})$ or 
$\SOL(\tau_{\mathrm{graph}(2)})$ logic.
We denote by $x,y,s,t,u,v,z$ the $\VARFO$ variables, by $A,B,F,S,U,W$ 
the $\VARSO$ variables and 
by $X,Y,Z$ the indeterminants in $\ind$.
For any formula $f$: 
\[\exists^k x(f(x))\defeq \exists x_1\cdots 
       \exists x_k(\bigwedge_{i\neq j} x_i\neq x_j \wedge 
    \bigwedge_{i=1}^k f(x_i)\:\wedge 
       \forall y((\bigwedge_{i=1}^k y\neq x_i)\rightarrow \neg f(y))). \]

For $D\subseteq A(G)$ and $S\subseteq E(G)$, $Touching(D,S)$ expresses the set 
of vertices or edges in $D$ which are adjacent to at least one edge from $S$,
$Cycle(S)$ is valid iff $S$ forms a cycle in $G$, and $Connected_S(u,v)$
expresses that $u$ is connected to $v$ through the edges in $S$.
These formulas take different form over vocabularies $\tau_{\mathrm{graph}(1)}$ 
and $\tau_{\mathrm{graph}(2)}$.
Over the vocabulary $\tau_{\mathrm{graph}(1)}$ $S$ is a symmetric relation,
and then:
\[ Touching(D,S) \defeq \{v:v\in D\wedge \exists u(S(v,u))\} \]
\[ Cycle(S) \defeq \forall u,v\in Touching(V,S)[\exists^2 y(S(u,y))\wedge Connected_S(u,v)] \]
\begin{eqnarray*}
  Connected_S(s,t) & \defeq & (s=t)\vee
     \exists U[U(s)\wedge U(t)\wedge 
       \forall x [U(x)\rightarrow \exists y(y\neq x\wedge S(x,y))]\wedge \\
   & & \neg\exists W[W(s)\wedge \neg W(t)\wedge\forall x[(W(x) \rightarrow \\
   & &   (U(x)\wedge \forall y((S(x,y)\wedge U(y)) \rightarrow W(y)))]].
\end{eqnarray*}
This formula expresses the fact that 
there is no subset $W \subsetneq U$ which contains $s$, does not contain
$t$, and such that for each vertex $x \in W$ all the neighbors of $x$ in $U$
are also on $W$
i.e., $W$ is a $S$-closed subset of $U$ which separates $s$ from $t$.

For the cases we use $\tau_{\mathrm{graph}(2)}$ ($A^G=V\cup E$),
we define shorthand formulas to identify an element 
of the universe to be an edge or a vertex respectively:
$P_E(x) \defeq \exists y(R(y,x)), ~P_V(x)\defeq x\in A\wedge \neg P_E(x)$,

Over the vocabulary $\tau_{\mathrm{graph}(2)}$ $S$ is a subset $S\subseteq \{x:P_E(x)\}$,
and then:
\[ Touching(D,S) \defeq \{x:x\in D\wedge \exists e[S(e)\wedge (N(x,e) \vee \exists u(N(u,e)\wedge N(u,x)))] \} \]
\[ Cycle(S) \defeq \forall u,v\in Touching(V,S)[\exists^2 e(S(e)\wedge N(u,e))\wedge Connected_S(u,v)] \]
\begin{eqnarray*}
   Connected_S(s,t) & \defeq & (s=t)\vee \exists U[\forall e(U(e) \rightarrow S(e))\wedge \\ 
     & & \forall v[((v=s \vee v=t)\rightarrow (U(s)\vee \exists^1 e(U(e)\wedge N(v,e)))) \wedge \\
     & &    ((P_V(v)\wedge v\neq s\wedge v\neq t) \rightarrow \\
     & &       (\neg\exists e(U(e)\wedge N(v,e))\vee
                  \exists^2 e(U(e)\wedge N(v,e))))].
\end{eqnarray*}
This formula expresses the fact that there is a subset $U\subseteq S$ which contains a direct
path from $s$ to $t$.

We also define $LastInComp(D,S)$ to be the set of elements in $D$ 
each of which is the last one by a given order $O$ in its component defined
by the edges in $S$. Formally:
\begin{equation}\label{eq:LastInComp}
LastInComp(D,S)\doteq \forall x\in D\forall y[(Connected_S(x,y)\wedge x\neq y) \rightarrow x\succ_O y].
\end{equation}

\begin{example}[Matching polynomial]
There are different versions of the matching polynomial discussed in
the literature (cf.~\cite{ar:HeilmannLieb72,bk:LovaszPlummer86,bk:GodsilRoyle01}),
for example \emph{matching generating polynomial}
$g(G,\lambda)=\sum_{i=0}^n a_i \lambda^i$ and \emph{matching defect
polynomial} $\mu(G,\lambda)=\sum_{i=0}^n (-1)^i a_i \lambda^{n-2i}$,
where $n=|V|$ and $a_i$ is the number of $i$-matchings in $G$. We
shall use the bivariate version that incorporates the both above:
\begin{equation} \label{general_matching_eqn}
M(G,X,Y)=\sum_{i=0}^n a_i X^{n-2i} Y^i
\end{equation}

Note that using the formulas defined above, if $F$ is a matching in $G$ then
$i=|F|$ and $n-2i=|V\setminus Touching(V,F)$.
This formula expressed as a $\SOL(\tau_{\mathrm{graph}(2)})$-polynomial expression is:
\begin{equation}\label{matching_SOL}
M(G,X,Y)=\sum_{F:Matching(F)}\:
            \left[\prod_{v:P_V(v)\wedge \neg(v\in Touching(V,F))} X\right]\:\cdot\:
            \left[\prod_{e:e\in F} Y\right]
\end{equation}
where 
\[ Matching(F)\defeq \forall e_1,e_2\in F[P_E(e_1)\wedge (e_1\neq e_2)\rightarrow 
                                            \neg\exists v(N(v,e_1)\wedge N(v,e_2))]. \]

\end{example}

\begin{example}[Tutte polynomial]
The classical two-variable Tutte polynomial
satisfies a subset expansion formula using spanning forests
(cf. for example B.Bollob\'as~\cite{bk:Bollobas99}).
Given a graph $G= \langle V \sqcup E, R \rangle$,
$O$ an ordering of $E$, and 
$F \subseteq E$ a spanning forest of $G$, i.e., 
each component of $(V,F)$ is a spanning tree of a
component of $G$.
An edge $e \in F$ is {\em  internally active (for $F, O$)}
if it is the first edge in the set
$ Cut_F(e)= $
$ \{ e' \in E : F -\{e\} \cup \{e'\} \mbox{ is a spanning forest} \} $.
An edge $e \in E- F$ is {\em  externally active (for $F, O$)}
if it is the first edge in the  unique cycle
$ Cycle_F(e) $ of $F \cup \{ e \}$.

For graphs $G$ with edge ordering $O$ the Tutte polynomial
satisfies
\begin{equation}
\label{tutte_spn}
T(G,X,Y) =
\sum_F X^i Y^j
\end{equation}
where the sum is over all spanning forests of $G$
and $i$ ($j$) is the number of internally (externally)
active edges of $F$ with respect to $O$.
Furthermore, this is independent of the ordering $O$.

%

Let $F\subset E(V)$ be a spanning forest of $G$,
i.e. $F$ contains no cycles and any connected component by $E(G)$ is also
connected by $F$:
\[ SpanningForest_G(F)\defeq \neg\exists U[U\subseteq F\wedge Cycle(U)] \wedge
     \forall v,u[Connected_E(v,u)\leftrightarrow Connected_F(v,u)] \]
The cycle of $e\not\in F$ is a set of edges $Z_F(e)$ such that:
\[e\in Z_F(e)\wedge (Z_F(e)\subseteq F\cup \{e\})\wedge Cycle(Z_F(e)).\]
The cut defined by $e\in F$ is a set of edges $U_F(e)$ such that:
\[U_F(e)\defeq \{e': SpanningForest_G((F\setminus \{e\})\cup\{e'\}) \}\]

Then, formula~\ref{tutte_spn} expressed as a $\SOL(\tau_{\mathrm{graph}(2)})$-polynomial expression is:
\begin{eqnarray}\label{tutte_SOL}
T(G,X,Y)=\sum_{F:SpanningForest_G(F)}\: &
            \left[\prod_{e:\forall e'((e'\in U_F(e)\wedge e\neq e')\rightarrow e\prec_O e')} X\right]\:\cdot\: \\
\nonumber
          & \left[\prod_{e:\forall e'((e'\in Z_F(e)\wedge e\neq e')\rightarrow e\prec_O e')} Y\right]
\end{eqnarray}

\end{example}

\begin{example}[The polynomial of the Pott's model]
\label{ex:potts}
This is a version of the Tutte polynomial used 
by A.Sokal \cite{ar:Sokal2005a}, known as the (bivariate)
{\em partition function}
of the  Pott's model:
\begin{equation}\label{sokal_eqn}
Z(G,q,v)=\sum_{A\subseteq E}q^{k(A)}v^{|A|}.
\end{equation}

Note that $k(A)=|LastInComp(V,A)|$.
Formula~\ref{sokal_eqn} expressed as a $\SOL(\tau_{\mathrm{graph}(2)})$-polynomial expression is:
\begin{equation}\label{sokal_SOL}
Z(G,q,v)=\sum_{A:A\subseteq E}\:
            \left[\prod_{v:v\in LastInComp(V,A)} q\right]\:\cdot\:
            \left[\prod_{e:e\in A} v\right].
\end{equation}

\end{example}


\subsection{Properties of $\SOL$-definable polynomials}

The following is taken 
from~\cite{ar:KotekMakowskyZilber08}.

\begin{proposition}
\label{pr:prop-s}
\ 
\begin{renumerate}
\item
If we write an $\SOL$-definable polynomial as a sum of monomials,
then the coefficients of the monomials are in $\N$.
\item
Let $M$ be an
$\SOL$-definable monomial viewed as a polynomial. 
Then $M$ is a product of a finite number
$s$ of terms of the form 
$\prod_{\bar{a}:\angl{\mathcal{M},\bar{a}}\models\phi_{i}}t_{i},$
where $i\in[s]$, $t_{i}\in\N\cup\ind$ and $\phi_{i}\in \SOL$. 
\item
The product of two 
$\SOL(\tau)$-definable polynomials
is again a
$\SOL(\tau)$-definable polynomial.
\item
The sum of two 
$\SOL(\tau)$-definable polynomials
is again a
$\SOL(\tau)$-definable polynomial.
\item\label{pro:collapsing-products}
Let $\Phi(\mathcal{A},\bar{X})$ be a $\SOL$-definable monomial and $P:Str(\tau)\to\N[\bar{X}]$ be~of~form 
\[ P(\mathcal{M},\bar{X})= 
\sum_{\bar{R}:\langle\mathcal{M},\bar{R}\rangle\models\chi_{R}}\,
\prod_{\bar{b}:\angl{\mathcal{M},\bar{R},\bar{b}}\models\psi}\,
\sum_{\bar{a}:\angl{\mathcal{M},\bar{R},\bar{a},\bar{b}}\models\phi} \Phi(\angl{\mathcal{M},R, \bar{a},\bar{b}},\bar{X}).
\]
Then $P(\mathcal{M},\bar{X})$ is a $\SOL$-definable polynomial.
\end{renumerate}
\end{proposition}
\subsection{Combinatorial polynomials}
\label{subse:combfun}

In the examples we need the fact that some combinatorial polynomials
are indeed $\SOL$-definable polynomials.
The question which combinatorial function can be written as $\SOL$-definable
polynomials is beyong the scope of this paper, and is the topic
of T. Kotek's thesis \cite{phd:Kotek}.

The following are all $\SOL$-definable polynomials. 
We denote by $\card{M}{v}{\varphi}$ the number of $\bar{v}$'s in $\mathcal{M}$ that satisfy $\varphi$.
\begin{description}

\item[Cardinality, I:] The cardinality of a definable
set 
$\card{M}{v}{\varphi}=\sum_{\bar{v}:\varphi(\bar{v})}1$ is an
evaluation of a $\SOL$-definable polynomial. 

\item[Cardinality, II:]
The cardinality as the exponent in a monomial 
\\
$X^{\card{M}{v}{\varphi}}=\prod_{\bar{v}:\varphi(\bar{v})}X$ is an 
$\SOL$-definable polynomial. 

\item[Factorials:] The factorial of the cardinality of a definable set
is an instance of a $\SOL$-definable polynomial:
\\
$\card{M}{v}{\varphi}! = \sum_{\pi:Func1to1(\pi,\{\bar{v}:\varphi(\bar{v})\},\{\bar{v}:\varphi(\bar{v})\})}1$,\\
where $Func1to1(\pi,A,B)$ says that $\pi$ is a one-to-one function from relation $A$ to relation $B$:
\begin{eqnarray*}
Func1to1(\pi,A,B) & \defeq & \forall \bar{v}\forall \bar{u}~[~\pi(\bar{v},\bar{u})\to 
     [\:\bar{v}\in A\:\wedge\: \bar{u}\in B\:\wedge\:  \\
       & & ~~~\neg\exists \bar{w}\:(\:
             (\bar{w}\neq \bar{v}\wedge \pi(\bar{w},\bar{u}))\vee (\bar{w}\neq \bar{u}\wedge \pi(\bar{v},\bar{w}))\:)]]. 
\end{eqnarray*}

%

\item[Falling factorial:]\label{pro:falling-factorial} 
The falling factorial
$$
(X)_{\card{M}{v}{\varphi}}= X \cdot (X-1) \cdot \ldots \cdot (X- \card{M}{v}{\varphi}
$$
is not an $\SOL$-definable polynomial, because it contains negative terms,
which contradicts Proposition \ref{pr:prop-s}.
However, if the underlying structure has a linear order,
then it is an evaluation of an $\SOL$-definable polynomial.
We write
$$
(X)_{\card{M}{v}{\varphi}}= 
\prod_{\bar{a}: \varphi} X - \card{M}{v}{\varphi_{< \bar{a}}}
$$
where $\varphi_{< \bar{a}}$ is the formula
$(\varphi(\bar{v}) \wedge \bar{v} < \bar{a})$ and $\bar{v} < \bar{a}$
is shorthand for the lexicographical order of tuples of vertices.
\end{description}

\section{Deconstruction of a signed graph and its valuation}
\label{se:rec-def}
In the following section we use the notation 
$\tau_{graph(1)}$ and
$\tau_{graph(2)}$
for
graph vocabularies as defined in Subsection \ref{se:solpol-examples}.
The definitions below are applicable for either vocabulary.

\subsection{Deconstruction trees}
Let $\tau \in \{\tau_{graph(1)}, \tau_{graph(2)}\}$.
\begin{definition}[Context]
Given a graph $G$ 
and $\vec{x}\in A^{m},~m \in \N$, 
a vector of elements of $G$, we call $\vec{x}$ an
$m$-\emph{context}. 
Given a vocabulary $\tau$ we denote by $\tau_m$ the vocabulary
$\tau$ augmented by $m$ constant symbols interpreted by
the $m$-context $\vec{x}$.
We denote by $\mathcal{G}_m$ the collection of graphs 
$\angl{G,\vec{x}}$
with an $m$-context.
\end{definition}

We now equip the graph 
$\angl{G,\vec{x}}$
with a linear ordering of its $m$-tuples.

\begin{definition}[Context ordering $\VALORD_m$]
Let $\tau_m^o = \tau \cup \{a_1, \ldots, a_m, O\}$
where the $a_i$'s are constants symbols and $O$ is a $2m$-ary relation symbol.
Let $\phi_{ord} \in \SOL(\tau_m^o)$.
The class $\VALORD_m$ consists of $\tau_m$-structures such that
\begin{renumerate}
\item
$\angl{G, \vec{x}, O} \in \VALORD$ iff $\angl{G, \vec{x}, O} \models \phi_{ord}$.
\item
The interpretation of $O$ is a linear ordering of the $m$-tuples of $G$.
\item
For every $\angl{G, \vec{x}}$  there is an $O \subset A^{2m}$ with
$\angl{G, \vec{x}, O} \models \phi_{ord}$.
\item
$\vec{x}$ is the first element in the ordering $O$ of
$\angl{G, \vec{x}, O}$ 
\end{renumerate}
We denote 
by $\bar{G}$ strutures of the form
$\angl{G, \vec{x}, O}$,  
by 
$A(\bar{G})$  the universe of $\bar{G}$, by
$R(\bar{G})$  the graph relation of $\bar{G}$, and by
$c(\bar{G})$  the context of $\bar{G}$, and by
$O(\bar{G})$  the context ordering of $\bar{G}$.
\end{definition}

\begin{definition}[$\SOL$-Deconstruction Scheme]\label{def:deconstruction-scheme}
Let $\Phi$ be a $\tau_m^o-\tau_m^o$-translation scheme. 
$\Phi$ is a $\SOL$-\emph{deconstruction scheme along $\VALORD$}, if
\begin{renumerate}
\item
$A^{\Phi^{\star}[\bar{G}]}\subsetneq A$;
\item
at least one element $x_i$ of $\vec{x}$ is deleted, 
i.e., $x_i \not\in A^{\Phi^{\star}[\bar{G}]}$;
\item
$O^{\Phi^{\star}[\bar{G}]} = O|_{A^{\Phi^{\star}[\bar{G}]}}$;
\item
If $\bar{G} \in \VALORD$ then $\Phi^{\star}[\bar{G}] \in \VALORD$;
\end{renumerate}
In this case we call $\Phi^{\star}$ a $\SOL$-\emph{deconstruction along $\VALORD$},
or simply a \emph{deconstruction}, if $\VALORD$ is clear from the context.
\end{definition}

\begin{definition}[Guarded $\SOL$-Deconstruction Scheme]\label{def:gdeconstruction-scheme}
A \emph{guarded $\SOL$-deconstruction} is a pair $(T, \varphi)$,
such that $T$ is a $\SOL$-deconstruction scheme and $\varphi$ is
a $\SOL(\tau_m^o)$-formula, and such that
$\Phi^{\star}(\bar{G})$ is a non-empty structure
for each $\bar{G}$ which satisfies $\varphi$.
\end{definition}

\begin{remark}
\ 
\begin{renumerate}
\item
Note that the formulas in $\Phi$ and the formula $\varphi$
may have up to $m$ additional
free individual variables for the $m$-context.
\item
We say that the guarded $\SOL$-deconstruction $(T, \varphi)$
is \emph{enabled} on a graph $\bar{G}$ 
if $\bar{G} \models \varphi$.
\item
One could have incorporated the guard in the definition of
$\Phi$, but this is not suitable here, because we want to
refer to the guard $\varphi$ explicitly.
\end{renumerate}
\end{remark}

A $\SOL$-\emph{deconstruction tree} for a graph $G$ with an $m$-context $\vec{x}$
and for a set of guarded deconstructions 
$\{(T_1, \varphi_1), \ldots , (T_{\ell}, \varphi_{\ell})\}$
is a tree each internal node of which is 
labeled by a graph with an $m$-context.
The arc from a node labeled with 
$\langle G_1, \vec{x}_1 \rangle$ 
to its child labeled with
$\langle G_2, \vec{x}_2 \rangle$ 
respectively,
is labeled with a guarded
deconstruction $(T_i, \varphi_i)$ such that 
$\langle G_1, \vec{x}_1 \rangle \models \varphi_i$ and
$G_2 = T_i^*[G_1, \vec{x}_1]$.
Additionally we require that
for each internal node labeled with 
$\langle G, \vec{x} \rangle$
and each guarded deconstruction  enabled on
$\langle G, \vec{x} \rangle$
there is an outgoing arc labeled by it.
Furthermore, each leaf of the deconstruction tree
is labeled by the empty graph.
With full noational details this looks as follows.

\begin{definition}[$\SOL$-Deconstruction tree along $\VALORD$]
Given a graph $\bar{G} \in \VALORD$ over $\tau_m^o$ and given
a set of guarded 
$\SOL$-definable deconstructions  schemes
$\{(T_i, \varphi_i)\},~(i=1,\ldots,l)$, 
we define a $\SOL$-\emph{deconstruction tree} $\Gamma=\Gamma(\bar{G})$ along $\VALORD$ as follows:
\begin{renumerate}
    \item We have $\ell$ partial functions $f_i, i \leq \ell$, denoting the $\ell$
          child relations.
    \item The root of $\Gamma$, $r$, is a node marked by $\bar{G}$. 
    \item Each internal node $n$ of $\Gamma$ is marked by a graph $\bar{G}_n$. 
    \item The child $f_i(n)$  of an internal node $n$ marked with a non-empty graph $\bar{G}_n$ is
          marked with 
          $T_i^{\star}(\bar{G}_n)$, 
          where
          $T_i^{\star}(\bar{G}_n)$ is enabled and not empty.
    \item If $T_i^{\star}(\bar{G}_n)$ is not enabled $f_i(n)$ is undefined.
    \item Each leaf in $\Gamma$ is marked by the empty graph.
\end{renumerate}
\end{definition}

With this definition we have

\begin{proposition}
For every set of guarded $\SOL$-deconstructions
$\mathcal{T} = \{(T_i, \varphi_i): i \leq \ell \}$
acting on  $\VALORD$ defined by $\varphi_{ord}$, and
for every $\bar{G} \in \VALORD$
there is at most one $\SOL$-deconstruction tree $\Gamma(\bar{x})$.
\end{proposition}

We denote by
$$\enabled(\vec{x})\defeq \bigvee_{i=1}^l \varphi_i(\vec{x})$$
and call the formula 
$\enabled(\vec{x})$
the \emph{deconstruction enabling} formula.
Note that the labeling of each internal node $n$ in the deconstruction tree 
must satisfy $ \bar{G}_n \models \enabled$.

The graph $\bar{G}_n$ associated with the node $n$ is called \emph{the world view of $n$}.
We denote the subtree of $\Gamma$ rooted at an internal node $n$
by $\Gamma_n= \Gamma_n(\bar{G}_n)$.

\subsection{The linear recurrence relation}

The recursive definition of a graph polynomial $P$
tells us how to compute $P(\bar{G})$ from $T_i^*(\bar{G})$.
The linear recurrence relation we have in mind takes the form
\begin{equation}
\label{def:graphpol-rec-defn}
\rec: 
P(\bar{G}) =
\sum_{i: \bar{G} \models \varphi_i}  \sigma_i(\bar{G}) \cdot P(T_i^*(\bar{G}))
\end{equation}
where
$\varphi_i$ is the guard of $T_i$.
We still have to specify what the coefficients $\sigma_i(\bar{G})$ 
are allowed to be.

\begin{definition}[Coefficients of the linear recurrence relation]
Let $\{\sigma_i: \VALORD \mapsto \ring\},~(i=1,\ldots,l)$
be a set of mappings such that each $\sigma_i$ 
is a map associated with $T_i$ which maps a graph with an $m$-context into
an element of $\ring$.
Furthermore we require that $\sigma_i(\bar{G})$ is
given by a short $\SOL$-polynomial expression.
\end{definition}

\subsection{Valuation of a deconstruction tree}
Given a deconstruction tree $\Gamma(G)$ 
we want to assign to $\Gamma(G)$ a value in $\ring$.

Given a graph $G$, a deconstruction tree $\Gamma(G)$ of $G$ 
and coefficients $\{\sigma_i\}$,
we compute the
deconstruction tree valuation by applying the formula below to
each internal node $n$ of $\Gamma(G)$:

\begin{equation}\label{eq:red_gen_form}
    P(\bar{G}_n)\defeq
              \sum_{
                \begin{array}{c}
                    i\in \{1,\ldots,l\} \\
                    s.t.~ \bar{G}_n \models \varphi_i
                \end{array}
              }
                \sigma_i(\bar{G}_n) 
                 \cdot P(T_i^\star(\bar{G_n}))
\end{equation}
If $n$ is a leaf
we define $P(G_n, \vec{x}_n) \defeq 1^{\ring}$.
This computation is well defined for every
ordered graph with a context $\bar{G}$, but the computation may depend
on the underlying order of the contexts.



\subsection{Well defined recursive definition}

\begin{definition}
A recursive definition of a graph polynomial $P$ is given by
a triple $(\mathcal{T}, \rec, \varphi_{ord})$, where
\begin{renumerate}
\item
$\mathcal{T} = \{(T_i, \varphi_i): i \leq \ell \}$
is a finite family of guarded $\SOL$-destruction schemes 
acting on  $\VALORD$ defined by $\varphi_{ord}$, and
\item
$$
\rec:
P(\bar{G}) =
\sum_{i: \bar{G} \models \varphi_i} 
\sigma_i(\bar{G}) \cdot P(T_i^*(\bar{G}))
$$
is a linear recurrence relation.
\end{renumerate}
\end{definition}

For the recursive definition 
$(\mathcal{T}, \rec, \varphi_{ord})$
of a graph polynomial $P$ to be well
defined we need several conditions to be satisfied.

\begin{definition}
A triple
$(\mathcal{T}, \rec, \varphi_{ord})$
is $\SOL$-{\emph feasible for} $P$ if the following conditions are
satisfied.
\begin{renumerate}
\item
$\VALORD$ is $\SOL$-definable by a $\SOL$-formula $\varphi_{ord}$.
\item
Every graph $G \in \mathcal{G}_m$ has an expansion $\bar{G}=\angl{G, \vec{x},O}$
with an order $O$ such that  
$\angl{G,\vec{x}, O} \models \varphi_{ord}$,
i.e., such that
$\angl{G,\vec{x}, O} \in \VALORD$.
\item
Every graph $\bar{G} \in \VALORD$ has a $\SOL$-deconstruction tree
$\Gamma(\bar{G})$.
\item
Given two orders $O_1$ and $O_2$ on $G$
and the corresponding deconstruction trees
$\Gamma(G,O_1), \Gamma(G,O_2)$ 
we have
$P(\Gamma(G,O_1))= P(\Gamma(G,O_2))$.
\end{renumerate}
\end{definition}

\begin{proposition}
Given a $\SOL$-feasible triple 
$(\mathcal{T}, \rec, \VALORD)$,
there is a unique graph invariant $P$
such that for all ordered graphs $\angl{G, O} \in \VALORD$
$$
P(G) = P(\Gamma(G,O))
$$
\end{proposition}

Note that we can replace the logic $\SOL$ in the definitions of this
section by other logics used in finite model theory, 
say Fixed Point Logic $\FPL$, Monadic Second Order Logic
$\MSOL$, etc. Such logics are defined in detail in, say \cite{bk:FMT}.
The choice of $\SOL$ here is a choice of convenience.
In Section \ref{se:conclu} we shall return to the use of other logics.

\subsection{Examples}
\label{se:deconstruct_examples}
\ \\

In all the examples below, the universe of $G$ is $A^G=V\cup E$,
the context is monadic ($m=1$) and we take $\VALORD_1$ to be defined 
by $\phi_{ord}\defeq \forall x,y[(P_E(x)\wedge P_V(y))\rightarrow x\prec_O y]$, 
i.e., we require the edges in $G$ to come before the 
vertices in the order $O$.

\begin{example}[Matching polynomial]
The bivariate matching polynomial
(cf. for example~\cite{ar:HeilmannLieb72,bk:LovaszPlummer86,bk:GodsilRoyle01})
is defined by 
$$
M(G,X,Y)=\sum_{i=0}^n a_i X^{n-2i} Y^i
$$
Alternatively, it can be also defined
by a linear recurrence relation as follows.
The initial conditions are
$M(E_1)=X$ and
$M(\emptyset)=1$. Additionally, it satisfies the recurrence relations
\begin{eqnarray}\label{rec_matching}
\nonumber M(G) &=& M(G_{-e}) + Y \cdot M(G_{\dagger e})\\
M(G_1 \oplus G_2) &=& M(G_1)\cdot M(G_2)
\end{eqnarray}
Here
$M(G_{\dagger e})$ is the graph obtained from $G$ by deleting
the edge $e=(u,v)$ together with the vertices $u$ and $v$
and all the edges incident with $u$ and $v$.

To express this defintion within our framework,
we take $A^G=V\cup E$
and $R=N\subseteq V\times E$ is the adjacency relation 
between vertices and edges.
We define shorthand formulas to identify an item 
of the universe to be edge or vertex respectively:
$P_E(x) \defeq \exists y(R(y,x)), ~P_V(x)\defeq x\in A\wedge \neg P_E(x)$,
and a formula which captures the universe elements which are 
removed during the extraction of an edge $x$:
\[ Extracted(x,y)\defeq [y=x\vee R(y,x)\vee \exists u(R(u,x)\wedge R(u,y))]. \]

The following table summarizes the formulas for the 
recursive definition of the matching polynomial.
\ \\
\ \\
\noindent
\begin{tabular}{|l|l|l|ll|l|}
  \hline
    & Action &                & $T_i[G,x]$  & $T_i[G,x]$    &               \\
  i & type   & $\varphi_i(x)$ & $\phi_i(y)$ & $\psi_i(y,z)$ & $\sigma_i(x)$ \\
  \hline
  1 & $G_{-v}$ & $P_V(x)\wedge\neg \exists y(R(x,y))$ & $y\neq x$ & $R(y,z)$ & $X$ \\
  2 & $G_{-e}$ & $P_E(x)$ & $y\neq x$ & $R(y,z)\wedge z\neq x$ & $1$ \\
  3 & $G_{\dagger e}$ & $P_E(x)$ & $\neg Extracted(x,y)$ 
      & $R(y,z)$ & $Y$ \\
  \hline
\end{tabular}
\ \\

Note that in this case, $\enabled(G,x)$ does not contain the case of $P_V(x)\wedge \exists y(R(x,y))$,
therefore not for every order $O$ there exists a valid fixed order deconstruction tree with order $O$.
However, any order $O$ in which all the edges come before all the vertices,
defines a valid fixed order deconstruction tree.
\end{example}

\begin{example}[Tutte polynomial]
The Tutte polynomial is defined (cf. for example~\cite{bk:Bollobas99,ar:BollobasRiordan99})
by the initial conditions
$T(E_1)=1$ and
$T(\emptyset)=1$
and has linear recurrence relation:
\begin{eqnarray}\label{rec_tutte}
\nonumber T(G,X,Y) & = & \left\{\begin{array}{lll}
                X \cdot T(G_{-e},X,Y) & if~e~is~a~bridge, \\
                Y \cdot T(G_{-e},X,Y) & if~e~is~a~loop, \\
                T(G_{/e},X,Y)+T(G_{-e},X,Y) & otherwise
                \end{array}
                \right.\\
T(G_1 \oplus G_2,X,Y)&=&T(G_1,X,Y)\cdot T(G_2,X,Y)
\end{eqnarray}
where a bridge is an edge removing which separates its connected 
component to two connected components.

As in the case of matching polynomial we define
$A^G=V\cup E$, $R=N\subseteq V\times E$,
$P_E(x)\defeq \exists y(R(y,x))$ and $P_V(x)\defeq x\in A\wedge \neg P_E(x)$ 
In addition we define the next shorthand formulas:\\
For any formula $f$: 
\[\exists^k x(f(x))\defeq \exists x_1\cdots \exists x_k(\bigwedge_{i\neq j} x_i\neq x_j \wedge 
    \bigwedge_{i=1}^k f(x_i)\:\wedge \forall y((\bigwedge_{i=1}^k y\neq x_i)\rightarrow \neg f(y))). \]
For any two monadic relations $U$ and $W$:
\[ U\subseteq W \equiv \forall x(U(x) \rightarrow W(x)) \]

We define formulas to express an edge being a bridge, a loop, or none of these, respectively:
\begin{eqnarray*}
\nonumber 
Bridge(x) & \defeq & P_E(x)\wedge \exists y,z[y\neq z\wedge R(y,x)\wedge R(z,x)\wedge \\
& &     \quad \neg \exists U(U\subseteq P_E \wedge \neg U(x) \wedge \exists u_1,u_2[ \\
& &     \qquad U(u_1)\wedge U(u_2)\wedge R(y,u_1)\wedge R(x,u_2)\wedge            \\ 
& &     \qquad \quad\forall u_3[(P_V(u_3)\wedge u_3\neq y\wedge u_3\neq z) \rightarrow \\
& &     \qquad \qquad (\neg \exists e_1(U(e_1)\wedge R(u_3,e_1)) \vee 
                         (\exists^2 e_2(U(e_2)\wedge R(u_3,e_2))))]])] \\
Loop(x) & \defeq & P_E(x) \wedge \exists^1 y(R(y,x))  \\
None(x) & \defeq & P_E(x) \wedge \neg Bridge(x) \wedge \neg Loop(x)
\end{eqnarray*}

In the case of contraction of edge $x$ we remove the edge and the smaller one (by order $O$)
of its end vertices $u,v$. The remaining end vertex $v$ becomes adjacent to all the edges 
which entered either of $u,v$. To describe this we need the next formulas:
\begin{eqnarray*}
\nonumber 
EdgeEnds(x,u,v) & \defeq & R(u,x)\wedge R(v,x)\wedge u\prec_O v    \\
Left(x,u)       & \defeq & P_E(x)\wedge \exists v(EdgeEnds(x,u,v)) \\
Right(x,v)      & \defeq & P_E(x)\wedge \exists u(EdgeEnds(x,u,v))
\end{eqnarray*}
The resulting adjacency relation is:
\[ \psi_{Contract}(x,y,z) \defeq  \exists u,v[EdgeEnds(x,u,v)\wedge (R(y,z)\vee (y=v\wedge R(u,z))] \]


The following table summarizes the formulas for the 
recursive definition of the Tutte polynomial.
\ \\
\ \\
\noindent
\begin{tabular}{|l|l|l|ll|l|}
  \hline
    & Action &                & $T_i[G,x]$  & $T_i[G,x]$    &               \\
  i & type   & $\varphi_i(x)$ & $\phi_i(y)$ & $\psi_i(y,z)$ & $\sigma_i(x)$ \\
  \hline
  1 & $G_{-e}$ & $Bridge(x)$ & $y\neq x$ & $R(y,z)$ & $X$ \\
  2 & $G_{-e}$ & $Loop(x)$ & $y\neq x$ & $R(y,z)$ & $Y$ \\
  3 & $G_{/e}$ & $None(x)$ & $\neg Left(x,y)$ & $\psi_{Contract}(x,y,z)$ & $1$ \\
  4 & $G_{-e}$ & $None(x)$ & $y\neq x $ & $R(y,z)$ & $1$ \\
  5 & $G_{-v}$ & $P_V(x)\wedge\neg \exists y(R(x,y))$ & $y\neq x$ & $R(y,z)$ & $1$ \\
  \hline
\end{tabular}
\ \\
\ \\
\end{example}

\begin{example}[Pott's model]
The polynomial $Z(G,q,v)$, called the Pott's model,
is defined (cf. for example~\cite{ar:Sokal2005a})
by the initial conditions
$Z(E_1)=q$ and
$Z(\emptyset)=1$,
and satisfies the linear recurrence relation
\begin{eqnarray}\label{rec_sokal}
\nonumber Z(G,q,v) & = & v \cdot Z(G_{/e},q,v) + Z(G_{-e},q,v)\\
Z(G_1 \sqcup G_2,q,v) & = & Z(G_1,q,v)\cdot Z(G_2,q,v)
\end{eqnarray}

Again we define $A^G=V\cup E$, $R=N\subseteq V\times E$,
$P_E(x)\defeq \exists y(R(y,x))$ and $P_V(x)\defeq x\in A\wedge \neg P_E(x)$ 
We also borrow the definition of $\psi_{Contract}(x,y,z)$ from the Tutte polynomial.

The following table summarizes the formulas for the 
recursive definition for the Pott's model.
\ \\
\ \\
\noindent
\begin{tabular}{|l|l|l|ll|l|}
  \hline
    & Action &                & $T_i[G,x]$  & $T_i[G,x]$    &               \\
  i & type   & $\varphi_i(x)$ & $\phi_i(y)$ & $\psi_i(y,z)$ & $\sigma_i(x)$ \\
  \hline
  1 & $G_{-v}$ & $P_V(x)$ & $P_V(x)\wedge\neg \exists y(R(x,y))$ & $R(y,z)$ & $q$ \\
  2 & $G_{/e}$ & $P_E(x)$ & $\neg Left(x,y)$ & $\psi_{Contract}(x,y,z)$ & $v$ \\
  3 & $G_{-e}$ & $P_E(x)$ & $y\neq x$ & $R(y,z)\wedge z\neq x$ & $1$ \\
  \hline
\end{tabular}
\ \\
\ \\
\end{example}




\section{Main result}
\label{se:main}

We now can state and prove our main result.

\begin{theorem}\label{th:main-result}
Let the triple $(\mathcal{T}, \rec, \varphi_{ord})$ be $\SOL$-feasible
defining a graph polynomial $P$.
Then
there exists a $\SOL$-polynomial expression $S$ such that 
for every $\bar{G} \models \varphi_{ord}$, and for every $z$,
$P(\Gamma(\bar{G})) = e(S,\bar{G},z)$.
\end{theorem}

\begin{figure}[ht!]
\begin{center}
\epsfig{file=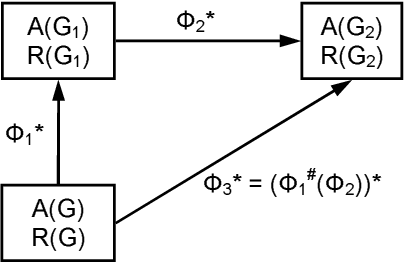,width=0.5\textwidth}
\end{center}
\caption{Translation scheme composition} \label{fig:trans-scheme-comp}
\end{figure}

\noindent
The following lemma, schematically represented by Figure~\ref{fig:trans-scheme-comp},
will be useful for the proof of the theorem:
\begin{lemma}\label{lem:trans-scheme-comp}
Let $\Phi_1 = \langle\phi_1,\psi_1\rangle$, $\Phi_2 = \langle\phi_2,\psi_2\rangle$ be translation schemes on graphs.
Let $G_1 = \Phi_1(G)$, $G_2 = \Phi_2(G_1)$, where $G, G_1, G_2$ are graphs over the same vocabulary.
Then there exists a translation scheme
$\Phi_3 = \Phi_1^{\sharp}(\Phi_2) = \langle \Phi_1^{\sharp}(\phi_2),\Phi_1^{\sharp}(\psi_2)\rangle $
such that $G_2 = \Phi_3(G)$.
\end{lemma}
\begin{Proof}
By definition of $\Phi_2$, we have
\begin{align*}
A(G_2) &= A^{\Phi_2^{\star}[G_1]} = \{a \in A(G_1): G_1 \models \phi_2(a)\}\\
R(G_2) &= R^{\Phi_2^{\star}[G_1]} = \{\vec{a} \in A(G_2)^2: G_1 \models \psi_2(\vec{a})\}
\end{align*}
By the fundamental property (Theorem~\ref{th:trans-funamental}), because $G_1 = \Phi_1^\star[G]$, we have
\begin{align*}
\forall a\in A(G_1) &(G_1 \models \phi_2(a) \leftrightarrow G \models [\Phi_1^\sharp(\phi_2)](a))\\
\forall \vec{a}\in A (G_1)^2 & (G_1 \models \psi_2(\vec{a}) \leftrightarrow G \models [\Phi_1^\sharp(\psi_2)](\vec{a}))
\end{align*}
This is equivalent to
\begin{align*}
\forall a\in A(G) &(G_1 \models (\phi_2(a) \land a\in A(G_1)) \leftrightarrow G \models [\Phi_1^\sharp(\phi_2)](a))\\
\forall \vec{a}\in A (G)^2 & (G_1 \models (\psi_2(\vec{a}) \land \vec{a}\in A(G_1)^2)
                               \leftrightarrow G \models [\Phi_1^\sharp(\psi_2)](\vec{a}))
\end{align*}
because if $A(G_1) \neq A(G)$ then $\Phi_1^\sharp$ relativizes $\phi_2,\psi_2$ to accept only $a \in A(G_1)$.
Thus we can take
$\Phi_3 = \Phi_1^{\sharp}(\Phi_2) = \langle \Phi_1^{\sharp}(\phi_2),\Phi_1^{\sharp}(\psi_2)\rangle $
\end{Proof}

\noindent
Now let us prove Theorem~\ref{th:main-result}.\\
\begin{Proof}
The proof is constructive.
The formula will simulate the iterative application of the reduction formula 
on some deconstruction tree $\Upsilon=\Upsilon(\bar{G})$.
The recursive definition $(\mathcal{T}, \rec, \varphi_{ord})$ is $\SOL$-feasible 
and therefore 
is invariant in the deconstruction tree, thus without loss of generality we can take $\Upsilon$
to be some fixed order deconstruction tree with a $\SOL$-feasible order $O$.
Note that the actual order of contexts in a branch $b$ is a sub-order $O_b$ of $O$.
A context $\vec{x} \in A^m$ might be omitted
from $O_b$ because the deconstructions performed along $b$ prior to the node marked by $\vec{x}$ might have deleted
an element of $\vec{x}$. This would make it impossible to use $\vec{x}$ as a context of any deconstruction.

The $\SOL$-polynomial expression we define, $S$, is a sum of the valuations of all the branches of $\Upsilon$.
Each branch $b$ is uniquely defined by the sequence of deconstructions ($T_i$-s) performed along the branch.
We define the vector of marks, $\vec{U} = (U_1, \ldots, U_l)$, which mark each context $\vec{x}$ according to
the deconstruction performed at the node of $\Upsilon$ marked by $\vec{x}$. Note that not all the contexts are
covered by $U_i$-s. Only the contexts that were not omitted from $O_b$ will be covered, as only at the nodes marked
by these contexts a deconstruction was performed. We mark the rest of the contexts by $D$.
Note also that the arity of each $U_i$, and of $D$, is $m$ - the cardinality of the contexts.

As follows from Definition~\ref{def:graphpol-rec-defn}, 
the valuation of the branch $b$ is the product of the
elementary valuations $\sigma_i(\vec{x})$ 
applied at each node $n$ marked by the context $\vec{x}$
such that $T_i^\star$ is applied at $n$, i.e., in our notation,
$$ \prod_{i=1}^l \prod_{\vec{x}:U_i(\vec{x})} \sigma_i(\vec{x}). $$
The $\SOL$-polynomial expression $S$ is now defined as follows:
\begin{equation}\label{eq:P_G_O}
S= 
\sum_{\vec{U},D:\Psi(\vec{U},D,O)} \prod_{i=1}^l \prod_{\vec{x}:U_i(\vec{x})} \sigma_i(\vec{x})
\end{equation}

Where $\Psi$ is
\begin{equation}\label{eq:psi-main}
\begin{split}
\Psi(\vec{U}&,D,O) = \\
     & \textrm{Disjoint}(\vec{U},D)\land \textrm{Cover}((D \cup \bigcup U_i), A^m)\land {} \\
     & \exists B\exists Q [ \\
     & \quad \forall \vec{x_0} [\:\textrm{first}_O(\vec{x_0})\rightarrow  \\
     & \qquad \forall u\forall v(B(\vec{x_0},u)\land
          (R(u,v)\leftrightarrow Q(\vec{x_0},u,v)) )\:] \land {}\\
     & \quad \forall \vec{x_1}\forall \vec{x_2}(\:\vec{x_2}=\textrm{next}_O(\vec{x_1})\rightarrow {} \\
     & \qquad \;\textrm{ChangeWorldView}(\vec{U},D,B,Q,\vec{x_1},\vec{x_2})\:) \; ]
\end{split}
\end{equation}

The predicate $\textrm{Disjoint}(\vec{U},D)$ means that the relations $U_1, \ldots, U_l, D$ are disjoint, and
$\textrm{Cover}((D \cup \bigcup U_i), A^m)$, meaning that each element of $A^m$ (i.e., each context) is marked either
by $D$ or by some $U_i$. We use $B \subseteq A^{m+1}$ and $Q \subseteq A^{m+2}$ to encode the world view of the nodes
of $\Upsilon$. Below we show that for a node $n$ on the branch $b$ which is marked by the context $\vec{x}$,
$B,Q$ satisfy $A(G_n)=\{v:B(\vec{x},v)\}$ and $R(G_n)=\{(v,u):Q(\vec{x},v,u)\}$. If a context $\vec{x}$ is the
first context in $O$ ($x_0$), then no deconstruction has been performed prior to the node marked by $x$.
Thus the world view of $x$ should be the original graph $G$. Otherwise, there exists a context $x_1$ which is an
immediate predecessor of $x$ in $O$. Then the world view of $x$ can be derived from the world view of $x_1$, and the
connection between these world views is described by the formula
$\textrm{ChangeWorldView}(\vec{U},D,B,Q,\vec{x_1},\vec{x})$

In order to define $\textrm{ChangeWorldView}$, the following definitions will be used:\\
For relations $B_1, Q_1$ such that $\rho(B_1)=1, \rho(Q_1)=2$ we define the translation scheme
$\Phi_{B_1,Q_1} = \langle B_1, Q_1 \rangle$.
For two relations $R_1, R_2$ of the same arity $\ell$ we overload the equality symbol to denote
$R_1 = R_2 \Leftrightarrow \forall u_1\ldots\forall u_l(R_1(u_1,\ldots, u_l) \leftrightarrow R_2(u_1,\ldots, u_l))$.\\

\begin{equation}\label{eq:change-world-view}
\begin{split}
  \textrm{Chan}&\textrm{geWorldView}(\vec{U},D,B,Q,\vec{x_1},\vec{x_2})= \\
     &   \exists B_1\exists B_2\exists Q_1\exists Q_2 [ \\
     &   \quad \forall u\forall v((B_1(u)\leftrightarrow B(\vec{x_1},u))\land {}
              (B_2(u)\leftrightarrow B(\vec{x_2},u))\land {}\\
     &   \qquad (Q_1(u,v)\leftrightarrow Q(\vec{x_1},u,v))\land {}
         (Q_2(u,v)\leftrightarrow Q(\vec{x_2},u,v)) )\land {}\\
     &   \quad \bigwedge_{i=1}^l ( U_i(\vec{x_1})\rightarrow [\:\Phi_{B_1,Q_1}^\sharp[\varphi_i](\vec{x_1})\land {} \\
     &   \qquad \qquad B_2=A^{\Phi_3^{\star}[G,\vec{x_1}]} \land {}\\
     &   \qquad \qquad Q_2=R^{\Phi_3^{\star}[G,\vec{x_1}]} \:] )\land {}\\
     &   \quad (D(\vec{x_1})\rightarrow [(\exists j \lnot B_1(\vec{x_1}[j]))\land B_1=B_2 \land Q_1=Q_2]) \qquad ]
\end{split}
\end{equation}
where $\rho(B_1)=\rho(B_2)=1,~\rho(Q_1)=\rho(Q_2)=2$ and $\Phi_3 = \Phi_{B_1,Q_1}^\sharp[T_i]$.

In accordance with the role of $B$ and $Q$, the first part of the formula defines the relations $B_i, Q_i$ to
comprise the world view of the context $x_i$.

The second part of the formula treats the case when the context
$\vec{x_1}$ is marked by some $U_i$, i.e., the case when the deconstruction $T_i^\star$ was applied at the node $n_1$
marked by $\vec{x_1}$. To make the application of $T_i^\star$ at $G_{n_1}$ possible,
$G_{n_1} \models \varphi_i(\vec{x_1})$ should hold. We need to find a formula $\widetilde{\varphi_i}$ such that
$G \models \widetilde{\varphi_i}(\vec{x_1})$ iff $G_{n_1} \models \varphi_i(\vec{x_1})$. $B_1, Q_1$ comprise
the world view of $x_1$, $G_{n_1}$. Thus by definition of $\Phi_{B_1,Q_1} = \langle B_1, Q_1 \rangle$, we have
that $\Phi_{B_1,Q_1}$ is a translation scheme translating $G$ to $G_{n_1}$.
Then, by Theorem~\ref{th:trans-funamental}, $G_{n_1} \models \varphi_i(\vec{x_1})$ iff
$G \models \Phi_{B_1,Q_1}^\sharp[\varphi_i](\vec{x_1})$, taking
$\widetilde{\varphi_i} \defeq \Phi_{B_1,Q_1}^\sharp[\varphi_i]$.

The world view of $x_2$, $G_{n_2}$, is the result of application of $T_i^\star$ to $G_{n_1}$,
and is comprised of $B_2, Q_2$.
Using Lemma~\ref{lem:trans-scheme-comp} applied to
$\Phi_1 \defeq \Phi_{B_1,Q_1}$ and $\Phi_2 = T_i$, we obtain that
$\Phi_{B_2,Q_2} = \Phi_{B_1,Q_1}^\sharp[T_i]$.

The last part of the formula treats the case when the context $\vec{x_1}$ (or part of it) is already deleted by
deconstructions applied to contexts which precede it in $O$. Therefore it should be marked by $D$. No deconstruction
is applied to $\vec{x_1}$, thus the world view of $\vec{x_1}$ and its successor, $\vec{x_2}$, are the same.
\end{Proof}

Note that if the coefficients $\sigma_i(\bar{G})$ of the recurrence relation are
given by short $\SOL$-polynomial expression  then the expression $S$ defines a
$\SOL$-polynomial.

\section{Derivations of subset expansion formulas}
\label{se:main_examples}


In this section we shall show how the proof of Theorem \ref{th:main-result}
can be applied to obtain a subset expansion formula
for the universal edge elimination polynomial~\cite{ar:AverbouchGodlinMakowsky08},
and the cover polynomial~\cite{ar:ChungGraham95}.

\subsection{The universal edge elimnation polynomial}
\label{se:use_main_ilia}
\ \\

The universal edge elimination polynomial $\xi(G,X,Y,Z)$ is a generalization 
of both the Matching and the Pott's model,
and is recursively defined in~\cite{ar:AverbouchGodlinMakowsky08}.

The initial conditions are
$\xi(E_1,X,Y,Z) = X$ and $\xi(\emptyset,X,Y,Z)=1$.

The recurrence relation is
\begin{eqnarray} \label{rec_AGM} 
\xi(G,X,Y,Z) & = & \xi(G_{-e},X,Y,Z) + y\cdot \xi(G_{/e},X,Y,Z) + z\cdot \xi(G_{\dagger e},X,Y,Z)\\
\nonumber
\xi(G_1 \oplus G_2,X,Y,Z) & = & \xi(G_1,X,Y,Z)\cdot \xi(G_2,X,Y,Z).
\end{eqnarray}

To express this defintion within our framework,
we define $A^G,~R,~P_E(x)$, $P_V(x)$,
$\psi_{Contract}(x,y,z)$ and $Extracted(x,y)$
similarly as in Example~\ref{ex:potts}.

\begin{table}
\caption{\label{tb:ilia-rec}
Formulas for the 
recursive definition of $\xi(G,X,Y,Z)$}
\begin{tabular}{|l|l|l|ll|l|}
  \hline
    & Action &                & $T_i[G,x]$  & $T_i[G,x]$    &               \\
  i & type   & $\varphi_i(x)$ & $\phi_i(y)$ & $\psi_i(y,z)$ & $\sigma_i(x)$ \\
  \hline
  1 & $G_{-v}$ & $P_V(x)$ & $P_V(x)\wedge\neg\exists y(R(x,y))$ & $R(y,z)$ & $X$ \\
  2 & $G_{-e}$ & $P_E(x)$ & $y\neq x$ & $R(y,z)\wedge z\neq x $ & $1$ \\
  3 & $G_{/e}$ & $P_E(x)$ & $\neg R(y,x)$ & $\psi_{Contract}(x,y,z)$ & $Y$ \\
  4 & $G_{\dagger e}$ & $P_E(x)$ & $\neg Extracted(x,y)$ & $R(y,z)$ & $Z$  \\
  \hline
\end{tabular}
\end{table}

Substituting the formulas of Table \ref{tb:ilia-rec} 
in the Equations (\ref{eq:P_G_O},\ref{eq:psi-main},\ref{eq:change-world-view})
we get a $\SOL$-polynomial expression.
This expression is a sum over the colorings $U_1,\ldots,U_4$ of $A^G$
of addends evaluated $\prod_{i=1}^4 \prod_{x:U_i(x)} \sigma_i(x)=X^{|U_1|}\cdot Y^{|U_3|}\cdot Z^{|U_4|}$.

Let $C$ be the set of the connected components of the graph $G_C\defeq (V(G), U_3\cup U_4)$.
In Formula (\ref{eq:change-world-view}), for each context $x_1$ satisfying $U_3(x_1)$
and $x_2=\textrm{next}_O(x_1)$ the contraction action on edge $x_1$ leaves one of 
its end verices. In other words, if $u,v\in V(G)$ and $\{(u,x_1),(v,x_1)\}\in R$
and $u\prec_O v$ then we have $B(x_1,u)\wedge B(x_1,v)$ but $\neg B(x_2,u)\wedge B(x_2,v)$.
Thus, action number 3 ($G_{/e}$) can not remove a whole connected component in $C$
from $\{y:B(x_2,y)\}$.

Therefore, for each component $c\in C$, actions 1 ($G_{-v}$) or 4 ($G_{\dagger e}$) 
must be used on the last vertex or edge in $c$ to eliminate whole of $c$ 
form $\{y:B(x,y)\}$ for some $x$ such that $U_1(x)$ or $U_4(x)$, respectively.

We divide the components in $C$ into two sets:
\begin{eqnarray*}
C_A & \defeq & \{ c\in C: \exists x\in c(U_1(x)) \} \\
C_B & \defeq & \{ c\in C: \exists x\in c(U_4(x)) \}
\end{eqnarray*}
and define the next edge sets:
\begin{eqnarray*}
A & \defeq & \{ x\in E(G):\exists c(x\in c\in C_A)\} \\
B & \defeq & \{ x\in E(G):\exists c(x\in c\in C_B)\}
\end{eqnarray*}

Recallin the definition of $Touching(D,S)$ and $LastInComp(D,S)$
from Section~\ref{se:solpol-examples} we get:
\begin{eqnarray*}
U_1 & = &  LastInComp(V,A\cup B)\setminus Touching(V,B)\} \\
U_3 & = &  A\cup B\setminus LastInComp(B,B) \\
U_4 & = &  LastInComp(B,B)
\end{eqnarray*}

If we rewrite Equation (\ref{eq:P_G_O}) using these terms, we get
the next simple $\SOL$-polynomial expression:
\begin{eqnarray}
\label{ilia_SOL}
\nonumber
\xi(G,X,Y,Z) & = & \sum_{A,B:A,B\subseteq E\wedge VertexDisjoint(A,B)}\:
                 \left[\prod_{v:v\in (LastInComp(V,A\cup B)\setminus Touching(V,B))} X\right]\:\cdot\: \\
             & & \left[\prod_{e:e\in (A\cup B\setminus LastInComp(B,B))} Y\right]\:\cdot\:
                 \left[\prod_{e:e\in LastInComp(B,B)} Z\right].
\end{eqnarray}
where $VertexDisjoint(A,B)\defeq \neg\exists v\exists a\in A\exists b\in B(N(v,a)\wedge N(v,b))$.

From this one can get
\begin{equation}
\label{exp_ERP}
\xi(G,X,Y,Z) = \sum_{(A \sqcup B) \subseteq E}
                 X^{k(A\sqcup B)-k_{cov}(B)}\cdot Y^{|A|+|B|-k_{cov}(B)}\cdot
                 Z^{k_{cov}(B)}
\end{equation}
where by abuse of notation we use $(A\sqcup B)\subseteq E$ for
summation over subsets $A,B \subseteq E$, such that the subsets of
vertices $V(A)$ and $V(B)$, covered by respective subset of edges,
are disjoint: $V(A) \cap V(B) = \emptyset$; $k(A)$ denotes the
number of spanning connected components in $(V,A)$, and $k_{cov}(B)$
denotes the number of covered connected components, i.e. the
connected components of $(V(B),B)$.

Note that $k(A\sqcup B)-k_{cov}(B)=|LastInComp(V,A\cup B)\setminus Touching(V,B)|$,
$|A|+|B|-k_{cov}(B)=|A\cup B\setminus LastInComp(B,B)|$ and
$k_{cov}(B)=|LastInComp(B,B)|$.

Now, Equation~\ref{exp_ERP} 
is the subset expansion formula for
$\xi(G,X,Y,Z)$ presented in~\cite{ar:AverbouchGodlinMakowsky08}.

\subsection{The cover polynomial}
\label{se:use_main_cover}
\ \\

The standard definition of the Cover polynomial for a directed
graph $D$ is (see~\cite{ar:ChungGraham95}):
\begin{eqnarray*}
\nonumber 
C(\emptyset) & = & 1, \\
C(E_n) & = & X^{\underline{n}}~\defeq~X(X-1)\cdots(X-n+1), \\
C(D)   & = & \left\{\begin{array}{lll}
                C(D_{-e})+C(D_{/e})        & if~e~is~a~loop, \\
                C(D_{-e})+Y\cdot C(D_{/e}) & if~e~is~a~not~a~loop
                \end{array} \right.
\end{eqnarray*}
where a contraction of a directed edge $e$ is defined in the following manner:
\begin{itemize}
\item 
If the edge is a loop then it and its adjacent vertex is deleted.
\item
Otherwise we remove this edge, replace both its adjacent vertices by a single vertex 
and keep all their adjecent edges which agree with the direction of $e$.
I.e., if $e=\angl{u,v}$ we remove them both, replace them by a new vertex $w$ and 
connect all edges $\angl{x,w}$ such that $\angl{x,u}\in E(D)$ and 
all edges $\angl{w,y}$ such that $\angl{v,y}\in E(D)$.
\end{itemize}

This polynomial is for directed graphs, we express the graph within an extended 
vocabulary $\tau_{\mathrm{direct-graph}(2)}=\angl{A,N^O,N^I}$ 
where the interprestation is: $A=V\cup E$ is the universe
of the graph, $N^O\subseteq V\times E$ is the adjacency relation for the outbound edges,
and $N^I\subseteq E\times V$ is the one for inbound edges.
The relevant shorthand formulas to identify an element 
of the universe to be an edge or a vertex respectively, are:
$P_E(x) \defeq \exists y,z[N^O(y,x)\wedge N^I(x,z)], ~P_V(x)\defeq x\in A\wedge \neg P_E(x)$.

Other shorthand formulas we use:
\begin{eqnarray*}
\nonumber 
DEdgeEnds(x,u,v) & \defeq & N^O(u,x)\wedge N^I(x,v) \\
DLoop(x) & \defeq & P_E(x) \wedge \exists y[N^O(y,x)\wedge N^I(x,y)]  \\
\psi^O_{Contract}(x,y,z) & \defeq & N^O(y,z) \\
\psi^I_{Contract}(x,y,z) & \defeq & \exists u,v[DEdgeEnds(x,u,v)\wedge(N^I(y,z)\vee (z=v\wedge N^I(y,u))] \\
DExtracted(x,y) & \defeq & \exists u,v[DEdgeEnds(x,u,v)\wedge y\neq u\wedge \neg N^O(u,y)\wedge \neg N^I(y,v)] \\
DLoopExtracted(x,y) & \defeq & \neg\exists u[N^O(u,x)\wedge (y=u \vee N^I(y,u)\vee N^O(u,y))]
\end{eqnarray*}


\begin{table}
\caption{
\label{tb:cover-rec}
Formulas for the recursive definition of the Cover polynomial}
\begin{tabular}{|l|l|l|l|}
  \hline
    & Action &                &               \\
  i & type   & $\varphi_i(x)$ & $\sigma_i(x)$ \\
  \hline
  1 & $D_{-e}$ & $P_E(x)$ & $1$ \\
  2 & $G_{/e}$ & $P_E(x)\wedge \neg DLoop(x)$ & $1$\\
  3 & $G_{/e}$ & $DLoop(x)$  & $Y$ \\
  4 & $G_{-v}$ & $\neg \exists y (P_E(y))$ & $X+(-1)^{\ring}\sum_{y:\phi_4(y)} 1^{\ring}$ \\
  \hline
\end{tabular}

\ \\
\ \\
\begin{tabular}{|l|l|lll|}
  \hline
    & Action & $T_i[G,x]$  & $T_i[G,x]$      & $T_i[G,x]$        \\
  i & type   & $\phi_i(y)$ & $\psi^O_i(y,z)$ & $\psi^I_i(y,z)$   \\
  \hline
  1 & $D_{-e}$ & $y\neq x$ & $N^O(y,z)\wedge z\neq x$ & $N^I(y,z)\wedge y\neq x$ \\
  2 & $G_{/e}$ & $DExtracted(x,y)$ & $\psi^O_{Contract}(x,y,z)$ & $\psi^I_{Contract}(x,y,z)$ \\
  3 & $G_{/e}$ & $DLoopExtracted(x,y)$ & $N^O(y,z)$ & $N^I(y,z)$  \\
  4 & $G_{-v}$ & $y\neq x$ & $\emptyset$ & $\emptyset$ \\
  \hline
\end{tabular}
\end{table}

Note that $\sigma_4(x)$ is a $\SOL$-definable polynomial so our main result validity
is supported by the last $\SOL$-definable polynomial property in Proposition~\ref{pr:prop-s}.

Substituting the formulas of Table \ref{tb:cover-rec} 
in the Equations (\ref{eq:P_G_O},\ref{eq:psi-main},\ref{eq:change-world-view})
we get a $\SOL(\tau_{\mathrm{direct-graph}(2)})$-polynomial expression.
Note that in this case Formula (\ref{eq:change-world-view}) should be 
extended to represent both the realtions $N^I$ and $N^O$.
This is peformed trivially by introducing $Q^I$ and $Q^O$ ternary relations
into Formulas (\ref{eq:psi-main}) and (\ref{eq:change-world-view}), instead
the single $Q$ relation.

This $\SOL(\tau_{\mathrm{direct-graph}(2)})$-polynomial expression 
is a sum over the colorings $U_1,\ldots,U_4$ of $A^G$
of addends evaluated $\prod_{i=1}^4 \prod_{x:U_i(x)} \sigma_i(x)$.

We use similar arguments as in previous section (\ref{se:use_main_ilia}).
Let $C$ be the connected components of $G_C\defeq (V(G), U_2\cup U_3)$.
To eliminate a component $c\in C$ from $\{y,B(x,y)\}$ for some context $y$
actions 3 ($G_{/e}$) or 4 ($G_{-v}$) must be used on the last edge 
or vertex of $c$.

We divide the components in $C$ into two sets:
\begin{eqnarray*}
C_P & \defeq & \{ c\in C: \exists x\in c(U_4(x)) \} \\
C_C & \defeq & \{ c\in C: \exists x\in c(U_3(x)) \}
\end{eqnarray*}

Note that if for edge $x_1$, such that $U_2(x_1)\vee U_3(x_1)$, we have $N^O(u,x_1)\wedge N^I(x_1,v)$,
then for $x_2=\textrm{next}_O(x_1)$ $\{y:B(x_2,y)\}$ does not  contain
any edges into $v$ or edges out of $u$.
Therefore, each vertex in $G_C\defeq (V(G), U_2\cup U_3)$ is adjecent to at most
one incoming and one outgoing edge.
Thus, each $c\in C$ are either a path or a cycle (a single vertex without a loop is a path
or it is a cycle if it has a loop).

Let $OnCycle(v,B)\defeq \exists U[U \subseteq B\wedge \exists e(U(e)\wedge N^O(v,e))\wedge Cycle(B)]$.
If we set $B\defeq \{x:U_2(x)\wedge U_3(x)\}$ then:
\begin{eqnarray}
\label{eq:U_cover}
\nonumber
U_3 & = & \{ e\in E:\exists c\in C_C (\{e\}=LastInComp(E,c))\} \\
U_4 & = & \{ v\in V:\exists c\in C_P (\{v\}=LastInComp(V,c))\} \\
    & = & \{ v\in LastInComp(V,B)\wedge OnCycle(v,B) \}
\end{eqnarray}
Note that by Equation~\ref{eq:U_cover} we have also $U_3=|\{ v\in LastInComp(V,B)\wedge OnCycle(v,B) \}|$.

Note that in this case we need to take the definitions of $LastInComp(V,A)$, $Cycle(B)$
and their subformulas with the relation $N$ replaced by $N^I$ or $N^O$ in accordance to
the context.

Because the context ordering $\VALORD_m$ permits only orders $O$ such that 
the vertices come after edges, for any choice of valid coloring $\vec{U}$ 
there exists a vertex $y$ such that its world view graph 
$\angl{B(y,\ldots),Q^I(y,\ldots),Q^O(y,\ldots)}=E_k$ for some $k$
and therefore for all $x\succ_O y$ we have $U_4(x)$ or $D(x)$.
For such vertices $x$ with $U_4(x)$, $\sigma_4(x)=X-k+1$
and in Formola~\ref{eq:P_G_O} we get $\prod_{x:U_i(x)} \sigma_i(x)=X^{\underline{|U_4|}}$.
Thus, $\prod_{i=1}^4 \prod_{x:U_i(x)} \sigma_i(x)=X^{\underline{|U_4|}}\cdot Y^{|U_3|}$.

We denote $CyclePathCover(B)$ to be valid iff for every vertex $v$ no two edges of $B$
emanate or enter $v$:
\begin{eqnarray*}
CyclePathCover(B) & \defeq & \forall v[P_V(v)\rightarrow 
     \neg\exists e_1,e_2(e_1\neq e_2\wedge \\
     & & [(N^O(v,e_1)\wedge N^O(v,e_2))\vee (N^I(e_1,v)\wedge N^I(e_2,v))])]
\end{eqnarray*}

If we rewrite Equation (\ref{eq:P_G_O}) using these terms, we get
the next simple
$\SOL(\tau_{\mathrm{direct-graph}(2)})$-polynomial expression:
\begin{equation}\label{cover_SOL}
C(D,X,Y)=\sum_{B,L:B\subseteq E\wedge L=LastInComp(V,B)}\:
            \left[(X)_{\{v:v\in L\wedge \neg OnCycle(v,B)\}}\right]\:\cdot\:
            \left[\prod_{v:v\in L\wedge OnCycle(v,B)} Y\right].
\end{equation}
where $(X)_{\{v:v\in L\wedge \neg OnCycle(v,B)\}}$ is a falling factorial
which by the properties listed in Section~\ref{pro:falling-factorial} 
is expressible by a $\SOL$-polynomial expression over $\ring$ which contains $\Z$.
Though Formula (\ref{cover_SOL}) is not a $\SOL$-polynomial expression in a normal form,
by Proposition ~\ref{pr:prop-s}, item (\ref{pro:collapsing-products}), it is still a $\SOL$-polynomial expression.

Formula (\ref{cover_SOL}) 
is equivalent to the one presented in~\cite{ar:ChungGraham95}:
\begin{equation}\label{cover_eqn}
C(D,X,Y)=\sum_{i,j} c_D(i,j) X^{\underline{i}} Y^j
\end{equation}
where $c_D(i,j)$ is the number of ways of covering all the vertices of $D$
with $i$ directed paths and $j$ directed cycles (all disjoint of each other),
$X^{\underline{i}}~\defeq~X(X-1)\cdots(X-i+1)$ and $X^{\underline{0}}\defeq 1$.
$c_D(i,j)$ is taken to be 0 when it is not defined, e.g., when $i<0$ or $j<0$.

\section{A graph polynomial with no recurrence relation}
\label{se:application}

In \cite{ar:NobleWelsh99} a graph polynomial 
$U(G,\bar{X}, Y)$ 
is introduced which generalises 
the Tutte polynomial, 
the matching polynomial, and
the stability polynomial.
$U(G,\bar{X},Y)$
is defined for a graph $G=(V,E)$ as
\begin{equation}
\label{eq:weighted}
U(G,\bar{X},Y)  =
\sum_{A \subseteq E}
y^{|A|-r(A)}
\prod_{i=1}^{|V|} X_i^{s(i,A)}
\end{equation}
where $s(i,A)$ denotes the number of connected components
of size $i$ in the spanning subgraph $(V,A)$,
and $r(A)= |V|- k(A)$ is the rank of $(V,A)$.

It is obtained from a graph polynomial 
$W_{G,w}(\bar{X},Y)$ 
for weighted graphs
$\angl{G, w}$ by setting all the weights equal $1$.
For the weighted version there is a recurrence relation reminiscent
of the one for the Tutte polynomial, but the edge contraction operation
for an edge $e=(v_1,v_2)$, wich results in a new vertex $u$,
gives  $u$
the weight 
$w(u)= w(v_1) + w(v_2)$.
For 
$W_{G,w}(\bar{X},Y)$ 
a subset expansion formula is proven, which is equivalent
to Equation (\ref{eq:weighted}), when all the weights are set to $1$. 
Equation (\ref{eq:weighted}) is used in
\cite{ar:NobleWelsh99} as the definition of the polynomial 
$U(G,\bar{X},Y)$  for graphs without weights.
It is noted that the recursive definition given for
$W_{G,w}(\bar{X},Y)$ does not work, as the edge contraction operation
for weighted graphs, when applied to the case where all weights
equal $1$, gives a graph with weight for the new vertex resulting from
the contraction.

We now show, that the polynomial
$U(G,\bar{X},Y)$   is not an $\SOL$-polynomial, and therefore
has no feasible recurrence relation in our sense.
To see this we note a simple property of $\SOL$-polynomials.

\begin{definition}
Let $\bar{X}=(X_1, \ldots, X_{n})$ be a set of variables, and
$$
P(G,\bar{X}) = 
\sum{\bar{A}} X_1^{f_1(G,\bar{A})} \cdot \ldots \cdot X_{n}^{f_{n}(G, \bar{A})}
$$
be a subset expansion of a graph polynomial $P$.
We say that $P$ is {\em invariant under variable renaming}
if for all graphs $G$ and for all permutation $\sigma: \N \rightarrow \N$
we have
$$
P(G, X_{\sigma(1)}, \ldots , X_{\sigma(n)}) =
\sum{\bar{A}} \prod_{i \leq n} X_{\sigma(i)}^{f_{\sigma(i)}(G, \bar{A})} 
$$
\end{definition}

The following is easy to see:
\begin{proposition}
Assume for
$$
P(G,\bar{X}) = 
\sum{\bar{A}} X_1^{f_1(G, \bar{A})} \cdot \ldots \cdot X_{n}^{f_{n}(G, \bar{A})}
$$
that for all $i \leq n$ the exponent $f_i(G, \bar{A})$ of $X_i$
is not dependent on $i$.
Then
$P(G,\bar{X})$ is
invariant under variable renaming.
In particular,
$\SOL$-polynomials are 
invariant under variable renaming.
\end{proposition}

\begin{proposition}
$U(G,\bar{X}, Y)$  is not invariant under
variable renaming.
\end{proposition}
\begin{proof}
Let $E_n$ be the graph consisting of $n$ isolated vertices.
Then $s(i,A) =  |A|$ if $i=1$ and
$s(i,A) =  0$ if $i \geq 2$. 
We have
$$
U(E_n, X_1, \ldots, X_n ,y)=
\sum_{A \subseteq E}
y^{|A|-r(A)}
\cdot X_1^{|A|}
$$
If we now set $\sigma(n)=n+1$ we get
$$
U(E_n, X_2, \ldots, X_{n+1} ,Y)=
\sum_{A \subseteq E}
y^{|A|-r(A)}
$$
\end{proof}

\begin{corollary}
\ 
\begin{renumerate}
\item
$U(G,\bar{X}, Y)$  is not a $\SOL$-definable polynomial.
\item
There is no feasible recursive definition of
$U(G,\bar{X}, Y)$. 
\end{renumerate}
\end{corollary}

\section{Conclusion and open problems}
\label{se:conclu}

We have shown with Theorem \ref{th:main-result} how to convert certain recursive definition
of  graph polynomials, the $\SOL$-feasible recursive definitions,
into $\SOL$-definable subset expansion formulas,
herewith generalizing many special cases from the literature,
in particular the classical results for the Tutte polynomial,
the interlace polynomial, and the matching polynomial.
We have also explained how Theorem \ref{th:main-result} was used in
\cite{ar:AverbouchGodlinMakowsky08} to
find a subset expansion formula for the universal edge elimination polynomial
$\xi(G,X,Y,Z)$.

Our framework does not cover all the graph
polynomials which appear in the literature.
We have not discussed graph polynomials where indeterminates
are indexed by elements of the graph.
This occurs for example in \cite{ar:Sokal2005a}.
Our framework can be easily adapted to this situation.
In this case renaming of the variables has to include also
a renaming of the elements of the universe.

The weighted graph polynomial from \cite{ar:NobleWelsh99},
however, is not invariant under variable renaming because
the integer index of the variables carries a graph theoretic meaning.
It is this feature which allows us to show that $U(G,\bar{X},Y)$
is not $\SOL$-definable.

We have not discussed the possibility of a converse of Theorem \ref{th:main-result}. 

\begin{oproblem}
Find a graph polynomial $P$ which is defined by a
$\SOL$-definable subset expansion formula and which is invaraint
under variable renaming, but which has no $\SOL$-feasible (linear)
recurrence relation.
\end{oproblem}

In our framework of $\SOL$-feasible recursive definitions
the recurrence relation is required to be \emph{linear}. 
We chose this restriction because we did
not want to generalize beyond the natural examples.

\begin{oproblem}
Are there combinatorially interesting graph polynomials
defined recursively by non-linear recurrence relations?
\end{oproblem}

\begin{oproblem}
Is there an analogue to Theorem \ref{th:main-result}
for non-linear recurrence relations?
\end{oproblem}

The choice of Second Order Logic $\SOL$ as the base logic 
for this approach is merely pragmatical.
It can be replaced by Fixed Point Logic $\FPL$ and extensions
of $\SOL$.
It seems not to work for Monadic Second Order Logic $\MSOL$.
In our proof of Theorem \ref{th:main-result} we have to quantify over
relations which are at least ternary, even if the recursive definition
is $\MSOL$-feasible.

\begin{oproblem}
Find a sufficent condition which ensures that an $\MSOL$-feasible
recursive definition can be converted into an $\MSOL$-definable
subset expansion formula.
\end{oproblem}


\end{document}